\documentclass[letterpaper]{article} 
\usepackage{aaai23}  
\usepackage{times}  
\usepackage{helvet}  
\usepackage{courier}  
\usepackage[hyphens]{url}  
\usepackage{graphicx} 
\usepackage{natbib}  
\usepackage{caption} 
\usepackage{hyperref}
\usepackage{algorithm}
\usepackage{algorithmic}
\usepackage{paralist}

\usepackage{multirow}
\usepackage{multicol}
\usepackage{amsmath}
\usepackage{color}
\usepackage{subfigure}
\usepackage{newfloat}
\usepackage{listings}

\DeclareCaptionStyle{ruled}{labelfont=normalfont,labelsep=colon,strut=off} 
\floatstyle{ruled}
\newfloat{listing}{tb}{lst}{}
\floatname{listing}{Listing}
\usepackage[switch]{lineno}

\nocopyright

\title{Cross Task Neural Architecture Search for EEG Signal Recognition}

\title{Cross Task Neural Architecture Search for EEG Signal Classifications}
\author {
    Yiqun Duan\textsuperscript{\rm 1}
    Zhen Wang\textsuperscript{\rm 2}
    Yi Li \textsuperscript{\rm 3}
    Jianhang Tang \textsuperscript{\rm 3}
    Yu-Kai Wang \textsuperscript{\rm 1}
    Chin-Teng Lin \textsuperscript{\rm 1}
}
\affiliations {
    \textsuperscript{\rm 1} CIBCI lab, Australian Artificial Intelligence Institute, School of Computer Science, University of Technology Sydney, Ultimo, NSW 2007, Australia\\
    \textsuperscript{\rm 2} School of Computer Science, The University of Sydney, Darlington, NSW 2008, Australia\\
    \textsuperscript{\rm 3} Department of Electronic and Computer Engineering, The Hong Kong University of Science and Technology, Australia\\
    yiqun.duan@student.uts.edu.au, zwan4121@uni.sydney.edu.au, ylini@connect.ust.hk, tangjh91@ysu.edu.cn, yukai.wang@uts.edu.au, chin-teng.lin@uts.edu.au, 
}

\usepackage{bibentry}
\begin{document}
\maketitle

\begin{abstract}
Electroencephalograms (EEGs) are brain dynamics measured outside of the brain, which have been widely utilized in non-invasive brain-computer interface applications. Recently, various neural network approaches have been proposed to improve the accuracy of EEG signal recognition. 
However, these approaches severely rely on manually designed network structures for different tasks which normally are not sharing the same empirical design cross-task-wise.  
In this paper, we propose a cross-task neural architecture search (CTNAS-EEG) framework for EEG signal recognition, which can automatically design the network structure across tasks and improve the recognition accuracy of EEG signals.
Specifically, a compatible search space for cross-task searching and an efficient constrained searching method is proposed to overcome challenges brought by EEG signals. By unifying structure search on different EEG tasks, this work is the first to explore and analyze the searched structure difference in cross-task-wise. Moreover, by introducing architecture search, this work is the first to analyze model performance by customizing model structure for each human subject. Detailed experimental results suggest that the proposed CTNAS-EEG could reach state-of-the-art performance on different EEG tasks, such as Motor Imagery (MI) and Emotion recognition. Extensive experiments and detailed analysis are provided as a good reference for follow-up researchers. We will release the source code~\footnote{code:  \href{https://github.com/DuanYiqun/CTNAS-EEG}{https://github.com/DuanYiqun/CTNAS-EEG}} to contribute to the open source community.


\end{abstract}


\section{Introduction}
\label{sec:1_Introduction}




Noninvasive Brain-Computer Interface (BCI), especially Electroencephalograms (EEGs) signal, is attracting increasing interest from researchers because of its safety and the convenience of recording. 
EEG signals measure electrical activity in the brain by using small metal discs (electrodes) attached to the scalp.
The precise recognition of EEG signals is the basis for most BCI applications, such as Motor Imaginary (MI)~\cite{saa2013discriminative,keerthi2021cnn}, Emotion~\cite{wang2014emotional,wang2021review}, Robotic Control~\cite{BCI_wheelchair}, and Gaming~\cite{BCI_gaming}.
Previous works are mostly {\textit{based on manually designed neural networks}}~\cite{EEGNet_paper,liu2016emotion} to recognize EEG signals. 
However, different applications {\textit{are not sharing the same neural model}}, designing the network architecture still requires a lot of expert knowledge and takes up much time in the EEG field. 
Therefore, we expect to introduce automatic neural network design to replace manual design and further improve the accuracy of EEG recognition for various applications.

Our approach is inspired by Neural Architecture Search (NAS) framework~\cite{DBLP:conf/iclr/ZophL17,DBLP:journals/corr/abs-1806-09055,chu2020darts,xu2020pcdarts}, which introduces automatic design of artificial neural networks. 
Yet, previous explorations are mostly in computer vision area. 
By utilizing NAS, we increase the {\textit{automation level}} into mostly manual designed structure {\textit{in EEG field}}.
Moreover, this could bring neural network {\textit{customization capability to each single human subject in the first time.}} 
However, adopting neural architecture search into EEG area meets critical {\textit{challenges due to unique properties of EEG signals:}}
1) EEG signals are time series data with spatial channel distribution with a low signal-to-noise ratio (SNR), which requires a different search space compared to existing NAS methods. 
2) The signal trails varies significantly between different human subjects~\cite{wierzgala2018most}. 
3) The required search space may vary severely between different recognition tasks (such as Emotion and Motor Imagery (MI)).

In this paper, we propose CTNAS-EEG (Cross Tasks Differentiable Architecture Search for EEG Signals), an efficient neural architecture search framework for general human brain dynamics, with the consideration of the listed challenges above. 
We propose a unique prototype network (Meta-Net) specially designed for EEG signals in Section~\ref{subsec: Metanet} to overcome challenge 1) listed above, which could preserve channel-wise information and time-wise information together for the search space to utilize.
Also, an efficient search space that is compatible with multiple tasks of EEG signals is proposed and described in Section~\ref{subsec: Searchspace} to overcome challenges 1) and 3).
For challenge 2) the subject difference, we follow strategies from previous works, where we first perform searching on a mixed subjects dataset to explore a unified structure. 
However, since we introduce architecture search into EEG signal processing, we are first able to customize structures for each subject and report the performance, which could also alleviate challenge 2). 
Considering the low SNR properties of EEG signals, 
we propose two structure constraints in Section~\ref{subsec: constraint} to improve the performance. 
First, we constrain the network scale with limited sizes, which could alleviate over-fitting by reducing redundant parameters and generate more slim structures for practical use. 
Second, the aggregated probability of selecting `skip connection' is constrained with a lower bound, which could provide an instant gradient backward path thus alleviating the training difficulty brought by low SNR.

We conduct detailed experiments in Section~\ref{sec:exp} and Appendix to illustrate the efficiency of our proposed CTNAS-EEG. 
The performance of CTNAS-EEG is compared with manually designed baselines on both MI (Section~\ref{subsec:performance MI}) and Emotion tasks (Section~\ref{subsec:performance Emtion}).
Since this work is the first to unify MI and Emotion tasks with NAS, we provide cross-task searching results analysis in Section~\ref{subsec:crosstask_op} as references for follow-up researchers. 
The main contributions of the CTNAS-EEG could be categorized fourfold:

\begin{compactitem}
    \item 
    CTNAS-EEG is the first approach that introduces state-of-the-art differentiable architecture search into EEG signal recognition and could unify multiple separated tasks. 
    \item 
    A novel Meta-Net, as well as a compatible search space, is first proposed especially designed for EEG signals.
    \item 
    Simple yet efficient constraints are first applied to EEG structures, which could alleviate over-fitting and ensure searched structures are practical to deploy.
    \item Experimental results suggest that the CTNAS-EEG could achieve competitive performance simultaneously on MI and Emotion datasets. Extensive studies provide a detailed analysis of search results. 
\end{compactitem}
\section{Related Works}

Both conventional approaches and deep learning approaches to recognize EEG signals have been investigated. 

\textbf{Motor Imaginary (MI):} The common spatial patterns (CSP)~\cite{CSP_original_paper} and filter bank common spatial pattern (FBCSP)~\cite{FBCSP} are first found effective to enhanced the performance of early machine learning methods.
Later work~\cite{hersche2018fast} has improved the accuracy by introducing a linear support vector machine (SVM) combined with Riemannian covariance matrices. 
As deep learning introduces better noise resistance, later deep learning methods could use both the pre-processed feature and the raw EEG signal as the input. 
EEGNet~\cite{EEGNet_paper}, Shallow ConvNet~\cite{Schirrmeister_EEG_CNN} and its variances are early works using raw input.
TPCT~\cite{li2019novel} combines advantages from both pre-processed feature-based input and raw signals by introducing large-scale CNN.
TCNet~\cite{ingolfsson2020eegtcnet} introduces temporal convolution and reaches a good balance between network scale and accuracy.

\textbf{Emotion:} Support vector machine~\cite{atkinson2016improving,ICDE19_SVM} and random forest~\cite{liu2016emotion} are first explored for two-category classification. Besides, \cite{tuncer2021new} proposed a fractal pattern feature extraction approach for emotion recognition. Multi-frequency bands~\cite{zheng2015investigating} are combined with deep belief networks (DBN).  
Spatial-temporal recurrent neural network (STRNN)~\cite{zhang2018spatial} for EEG-based emotion detection, where they also split the EEG signals into five frequency bands for extracting DE features. Besides, \cite{krishna2019efficient} proposed a unique mixture model, which is well-performed on EEG signals interfered by noises based on asymmetric distribution.

\textbf{Neural Architecture Search}
Neural Architecture Search (NAS) has attracted extensive attention for its potential to design efficient networks automatically \cite{liu2019auto,ghiasi2019fpn,li2020neural,gao2021global2local}. 
Earlier searching works are based on reinforcement learning \cite{zoph2016neural,zoph2018learning} and evolutionary algorithms \cite{xie2017genetic,real2019regularized}, where most of them require `training-from-scratch' on various candidates and select candidates on each step by trained meta-controller (reinforcement learning) or crossover evolution.
In order to reduce the searching cost, more recent works such as one-shot NAS methods~\cite{guo2019single,chu2019fairnas} and gradient-based methods \cite{liu2018darts,he2020milenas} utilized the weight sharing strategy, where the searched results are presented as \textit{subnet} of a \textit{MetaNet}. 
Basically, the searching performs weights optimization on the \textit{MetaNet} which contains all possible operators and seeks a feasible structure hyper-parameters which decides the \textit{subnet}, the searched result.
We design our structure search methods for EEG signals based on gradient-based methods as it requires lower computation cost~\cite{jin2019rc}. 
An early attempt~\cite{rapaport2019eegnas} has explored simple scenarios of introducing NAS with EEG. 
But the search space is still the computer vision search space, yet the experiments are merely on motor imaginary and could not reach SOTA performance.

\begin{figure*}[hbpt]{
\centering
\subfigure[MeataNet]{
\includegraphics[width=0.8\textwidth]{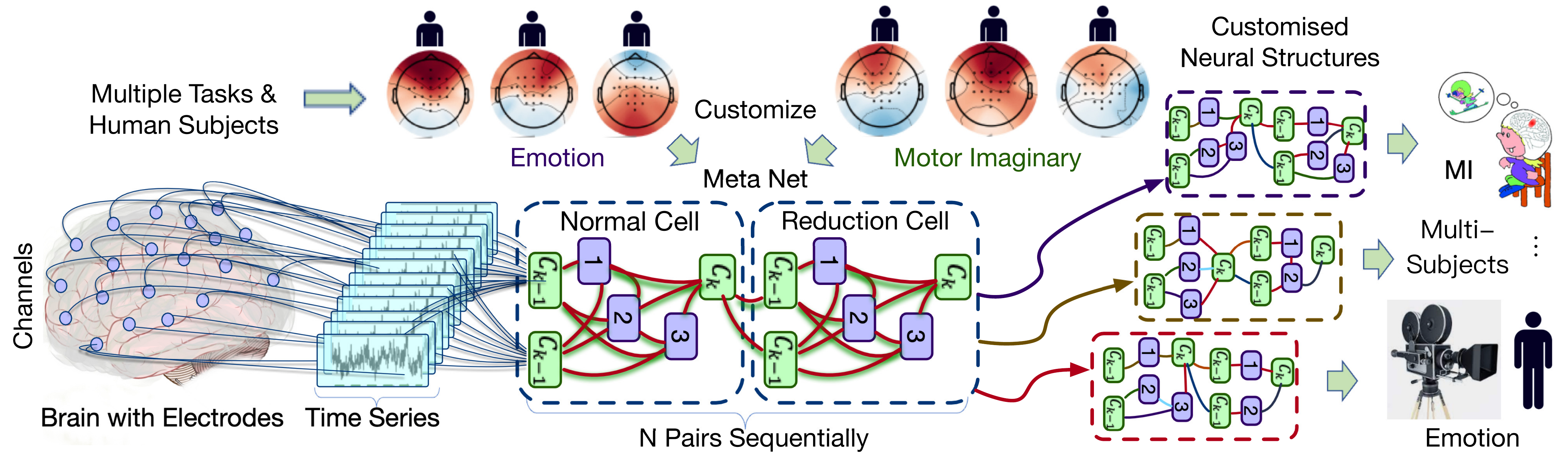}\label{fig:meta-net}
}
\quad
\subfigure[Search Space]{
\includegraphics[width=0.155\textwidth]{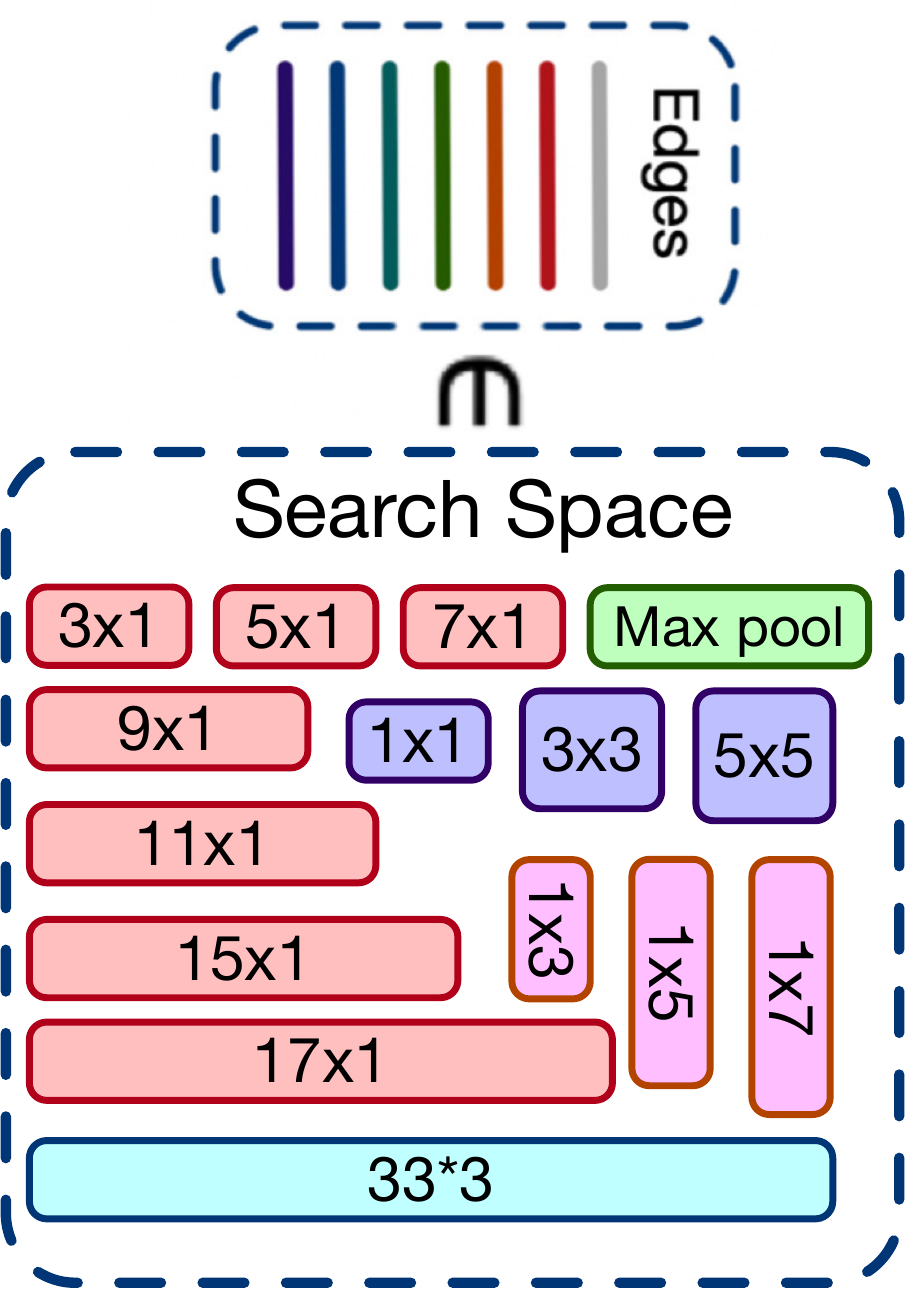}\label{fig:operator_space}
}

\caption{\label{fig:metanetwith_searchspace} 
The overall paradigm structure of CTNAS-EEG, where (a) illustrate how CTNAS receives brain dynamics from human brain and customize neural model not merely for each task but also each human subject. 
Content (b) denotes the search space where the weight of search space shares for each EEG channel. 
}}
\end{figure*}

\section{Cross Task Search for EEG Signal}
\label{sec:methods}
In this section, we will provide technical details of the proposed CTNAS-EEG\footnote{A brief preliminary background knowledge of differentiable architecture search (DARTS) is provided in Appendix~\ref{subsec: Preliminaries}. }.
We first give an overall introduction to the Meta-Net and feature design in Section~\ref{subsec: Metanet} 
An effective and compatible search space for EEG signals is introduced in Section~\ref{subsec: Searchspace}. 
Furthermore, we design structure constraint and solve the constrained optimizing problem in Section~\ref{subsec: constraint}

\subsection{Meta-Net and Stem Feature Processing}
\label{subsec: Metanet}

One significant property of EEG signals is that it contains both channel-wise\footnote{Each channel of the EEG signals has physical meanings, where it denotes electrode sensory data in each brain area.} information and time-wise information. 
The brain waves can present significant feature only if it is accumulated to a certain time-wise interval.
%
Previous methods may have introduced strong human priors, such as common spatial patterns, power spectral density, and differential entropy. 
Since our intuition is to \textbf{\textit{let the machine decide}} the processing structure for signals, this work directly uses \textbf{raw EEG signal} as input. 
Given EEG signal trails with shape $channel\times samplepoints$, we first slice and stack the whole data with a sliding window with overlap as shown in Figure~\ref{fig:metanetwith_searchspace}. After that, the feature shape becomes $bs \times channels \times slices \times samplepoints$. 
After that, a global channel-wise 1x1 convolution layer is applied to mix channel-wise information from EEG. Since the EEG channels have physical meanings, the MetaNet does not perform any channel reduction throughout the network forward from then. 
There are two kinds of searchable cells in the MetaNet, the Normal Cell, and the Reduction Cell. 
The Reduction Cell is in charge of feature downsampling after certain operators, while the normal cell only contains a combination of operators.
Several Normal Cell and Reduction Cell pairs connect sequentially to form the network structure. 
After the features are processed by N pairs of search cells, the features are fed into an FC layer to output classification logits. 


\subsection{Search Space}
\label{subsec: Searchspace}
As mentioned in Appendix~\ref{subsec: Preliminaries} and Equation~\ref{eq:mixedop}, during the search period, the search space provides a set of potential operators to be selected by each edge. Please refer to Appendix~\ref{ap:searchoperators} for detailed operators.
\subsubsection{Operators}
\label{subsubsec:op}
The output of each edge inside search cells is decided by the search space, as the output is weighted mixture~\footnote{Please refer to Appendix Section~\ref{subsec: Preliminaries} for detailed backgrounds. } of all the operators in search space during the training period.  
In that case, efficient and compatible are important at the same time.
If the search space contains many inefficient operators, the `bad' operators will bring negative effects to the descending of the MetaNet.
Meanwhile, if a search space is not compatible for multiple tasks, the searching algorithm cannot select proper operators as it does not exist in the search space. 
A typical sampling approach of the raw signals collected via 10-20 system~\cite{herwig2003using} has a shape with $channels*samplepoints$, where the sample points are generated from the sampling rate in a certain period. 
In order to aggregate more time period, CTNAS-EEG proposed to utilize more `narrow' operators, such as $Conv 5*1, Conv 11*1$ or even $Conv 17*1$. 
Meanwhile, we also put classic square convolutional kernels such as $Conv 3*3$ to provide spatial extraction ability.
Thin operators such as $Conv 1*3, Conv 1*5$ are also provided in case the MetaNet needs operation between slices. 
For each shape, we provide a `dilation' and `separable' version of the operators in the search space to further enlarge the perception field. 
Please refer to Appendix~\ref{ap:searchoperators} for detailed operator configurations. 

\subsubsection{Pooling}
Pooling operation is important in most deep neural networks~\cite{boureau2010theoretical,duan2019learning,CVPR22_Transformer} as it could 1) down-sampling feature map to reduce computational consumption, 2) reduce over-fitting by aggregating spatial nearly distributed feature value. 
Especially, the second point becomes more important to deal with plenty of noise in raw EEG signals.
According to our extensive experiments, average pooling is simply not working well. 
In that case, we put max pooling with various shapes in the search space. 

\subsubsection{Activation}
The activation function provides nonlinearity and improves the representation ability of a network. 
We add an activation layer after every operator in the search space. 
Extensive experiments suggest that the traditional Relu and Sigmoid function is not performing well. Instead, we use Elu and LeakyRelu~\cite{nwankpa2018activation} in our search space. This also fits observations from previous works in EEG signal recognition.

\subsection{Constraint Optimizing}
\label{subsec: constraint}

As mentioned in introduction that we need to 1) limit the searched network scale for practical usage, and 2) improve the sampling probability of skip connection for the low signal-to-noise-ration properties of EEG signals. 
In that case, CTNAS-EEG introduces structure constraints to the searching process to reach our goal.
Different from the original DARTS algorithm mentioned in Appendix \ref{subsec: Preliminaries} Preliminaries, we form our optimization objective function with structure constraints. 
And the structure constraint is optimized through relaxation of the objective function using Augmented Lagrangian. 
Please refer to Appendix~\ref{subsec:hyperparam} for extensive experiments on the proposed structure constraint. 

\textit{The two proposed structure constraint is defined respectively as below:}
1) 
The network parameter scale $\mathbf{\Omega}(\theta)$ is constrained within a certain range. 
2) Also we constraint the network to increase the probability of using skip connection $\mathbf{\Phi}(\theta)$ to make the whole MetaNet with a more instant gradient backward path. 
The instant path could increase the noise-resistant ability thus stabilizing the searching process given EEG raw signals with low SNR.
The optimization problem is defined in Equation~\ref{eq:constraintproblem}:
\begin{align}
\label{eq:constraintproblem} 
    \nonumber &\mathop{\rm min}\limits_{\theta} \mathcal{L}_{\rm val}(w^{*}(\theta),\theta) \\
    s.t. \quad &w^{*}(\theta)={\rm arg min}_{w} \mathcal{L}_{\rm train}(w(\theta))\\ \nonumber
    \quad & \mathbf{C}_{l} < \mathbf{\Omega}(\theta)< \mathbf{C}_{h}, \\
    \nonumber \quad & \mathbf{\Phi}(\theta) > P_{skip}(t), P_{skip}(t)= \beta {e}^{-t},
\end{align} 
where $\mathbf{\Omega}(\theta)$ denotes the total parameter scale given structure parameter $\theta$, $\mathbf{C}_{l}$ and $\mathbf{C}_{h}$ respectively denotes the lower and upper bound correspond to current structure parameters. 
$\mathbf{\Phi}(\theta)$ denotes the probability indicator of selecting a skip connection, we force the network to select more skip connections in earlier stages of training. 
The lower bound of the probability $P_{skip}(t)=\beta {e}^{-t}$ will decay to zero as the training time $t$ increasing.
Yet, the parameter scale is discrete if we directly count the parameters related to the current operator. 
In that case, we apply a similar relaxation with the original DARTS~\cite{DBLP:journals/corr/abs-1806-09055} that we relax the discrete network parameter scales to a continuous function $\mathbf{\Omega}(\cdot)$ of the SoftMax value of the structure $\theta$ as it defined in Equation~\ref{eq:structure_descriptor}:
\begin{equation}
\label{eq:structure_descriptor} 
\mathbf{\Omega}(\theta) = \sum_{i<j} 
\frac{\sigma(\theta_{o}^{(i,j)}) {\rm exp}(\theta_{o}^{(i,j)})}{ \sum_{o^{'} \in O} {\rm exp}( \theta_{o^{'}}^{(i,j)})}, 
\end{equation} 
where the $\mathbf{\sigma}$ is a matrix that returns the normalized resources costs of operators in the search space given the weight of operators $\theta_{(i,j)}$ between node $i$ and node $j$. 
Matrix $\mathbf{\sigma}$ can be determined by giving the resource cost of each operator could be calculated from its weights, where the value of resource cost is between 0 to 1 by normalizing the weights between all weights of the operators in the search space.
Similarly, we define the skip connection indicator as shown in Equation~\ref{eq:skip_descriptor}.
\begin{equation}
\label{eq:skip_descriptor} 
\mathbf{\Phi}(\theta) = \sum_{i<j} 
    \frac{ {\rm exp}(\theta_{o=skip}^{(i,j)}/T)}{ \sum_{o^{'} \in O} {\rm exp}( \theta_{o^{'}}^{(i,j)}/T)}, 
\end{equation} 
where $\theta_{o=skip}^{(i,j)}$ denotes that we only count probabilities of skip connection to calculate $\mathbf{\Phi}(\theta)$. T is the temperature parameter widely used in knowledge distillation~\cite{gou2021knowledge,ICML20_Distillation}, which could make the output of the Softmax distributed more evenly, thus avoiding severe structure fluctuation at an earlier stage. 
Since the constrained problem defined in Equation~\ref{eq:constraintproblem} is not strict convex for optimization, CTNAS-EEG basically follows the effective relaxation of current DARTS papers, which alternatively update the weights $w$ and the structure parameter $\theta$. 
Meanwhile, we put the structure constraint into consideration while updating the structure parameter $\theta$. 
The constraint optimization problem is solved by transferring constraint to penalty item in objective function via Augmented Lagrangian~\cite{boyd2004convex}. The transformed objective function is defined in Equation~\ref{eq:lagarangian_sc}:
\begin{equation}
\label{eq:lagarangian_sc} 
\resizebox{.9\hsize}{!}{$
\begin{split}
\mathcal{L}_{lag_{\theta}} &= \mathcal{L}_{val}(w^{*}(\theta),\theta) + \lambda_{1}{\rm max}((\mathbf{C}_{l}-\lambda_{2}\mathbf{\Omega}(\theta)),0)\\ + \lambda_{2} &{\rm max}((\mathbf{\Omega}(\theta)-\mathbf{C}_{l}),0)+ \lambda_{3}{\rm max}((\mathbf{\Phi}(\theta)-\beta e^{-t}),0),
\end{split}
$} 
\end{equation}
where $\lambda_{1}$ and $\lambda_{2}$ are respectively the weight of the lower and upper bound of the network structure constraint. $\lambda_{3}$ is the weight of skip connection constraint. By optimizing the objective function defined in Equation~\ref{eq:lagarangian_sc}, the structure parameter $\theta$ could be updated while keeping the structure constrained. 
The overall iteration process could be summarized in Algorithm~\ref{algo:1}. 

\begin{algorithm}[hpbt]

\caption{Training Procedure of CTNAS-EEGß}
\label{algo:1}  
\begin{algorithmic}[1]
\REQUIRE Set of mixed operators $o_(i,j)$ parameterized by $\theta_(i,j)$ for each edge $(i, j)$.

\WHILE {Step $<$ Max and Not Converged}
\STATE 1. Fix current $\theta$ as optimal $\theta^{*}$, obtain optimal $w^{*}$ by descending along $\nabla \mathcal{L}_{\rm train}(w({\theta}^{*}))$
\STATE 2. Fix current $w$ as optimal $w^{*}$, obtain optimal $\theta^{*}$ by descending along $\nabla \mathcal{L}_{lag_{\theta}}$ as defined in Equation~\ref{eq:lagarangian_sc}
\ENDWHILE

\noindent Derive the final architecture based on the learned $\theta$.
\label{code:recentEnd}
\end{algorithmic}
\end{algorithm}

\section{Experiments}
\label{sec:exp}

This section presents experiments to illustrate the effectiveness of the proposed CTNAS-EEG framework. 
We give the description of the datasets and the experimental settings in Section~\ref{subsec:dataset} and Section~\ref{subsec:esetting}. 
The CTNAS-EEG performance is verified in customizing networks for two important tasks, MI and Emotion, reported in Section~\ref{subsec:performance MI} and Section~\ref{subsec:performance Emtion}, respectively. 
Due to limited paper length, please refer to Section~\ref{subsec:hyperparam} for a detailed ablation study. 
Experimental results support that the proposed CTNAS-EEG could reaches SOTA performance on both cross-subject and within-subject tasks.

\subsection{Datasets}
\label{subsec:dataset}

We compare the model performance on the datasets of two most important tasks, MI and Emotion, in EEG area.
\subsubsection{BCI Competition IV for MI}
For MI, we select BCI Competition IV dataset~\cite{tangermann2012review} 2a to conduct our experiments.
Both datasets contain EEG and EOG signals with a sampling frequency of 250 Hz from nine subjects. 
For dataset 2a, subjects were required to perform four classes (left hand, right hand, feet, and tongue) MI classification. 
The BCI-competition IV dataset is collected under a widely-used 10-20 system, where 22 EEG channels and 3 EOG channels are provided.
In this experiment, we only consider the 3 seconds `Imaginary period'~\footnote{The BCI Competition IV~\cite{tangermann2012review} dataset is collected through a four-period paradigm containing `Fixation Cross', `Cue', `Imaginary Period', and `Pause'.} which contains 750 sample points in each trail.  
The sample distribution between different labels is balanced naturally while the data is collected.

\subsubsection{SEED Dataset for Emotion}
SEED-IV dataset contains 15 subjects, and each subject has three sessions. It includes four emotion types `happy', `neutral', `sad', and `fear', and each emotion has 6 film clips. Thus there are a total of 24 trails, and each trail has 12-64 samples for one session of each subject. Then there are
about 830 samples in one session;
SEED-V dataset also contains 15 subjects, each with three sessions. However, the emotion types have increased to 6 with `disgust', `Fear', `Sad', `Neutral', `Happy'. 
SEED-V dataset is a multi-modal Emotion dataset, yet, we only use the EEG signals from it. 
The SEED datasets are recorded with a sampling rate 1000Hz with 62 channels.

\subsection{Experimental Setting}
\label{subsec:esetting}

\subsubsection{Data Processing}
Since the purpose is to unify the separate MI and Emotion tasks for the same searching framework, we re-sample all the datapoint from a different dataset with the same sampling rate of 250hz. 
In that case, the search results are comparable as all the experiments are searching on the same time scale. 
To analyze more results from the searching process, the logic is to remain raw EEG signals unchanged except for necessary normalization and filter. 
We largely keep the data processing details the same with EEGNet~\cite{EEGNet_paper}, where we only apply a cross-domain value normalization and use the normalized raw input as the model input. 
The slicing window size is 400 with stride step size 50, which means we use overlapped raw signal input.
We also do not apply data reproduction on the numbers of the trials, and only the original trials are used for our experiments.

\subsubsection{Implementation Details}
The whole search algorithm is implemented in PyTorch\footnote{\href{https://pytorch.org/}{https://pytorch.org/}}, a convenient deep learning library based on python.
Due to limited page length, please refer to Appendix~\ref{ap:implementation} for more technical details. 
We will release the code on Github to contribute to the open-source community for reproduction usage.

\subsubsection{Evaluation Metrics}

The evaluation metrics used in this framework are normally the same as previous methods in the EEG signals classification, including subject classification accuracy and subject-dependent classification accuracy.

\subsection{Motor Imaginary Performance}
\label{subsec:performance MI}
\subsubsection{Mixed Subjects Performance}
\label{subsec:mixedsubjects}

We first evaluate our proposal on mixed subject performance, where we mix the training $\&$ validation trails from all 9 subjects together into the integrated training  $\&$ validation set. 
By comparing our proposal with the previous state-of-the-art methods on the classification task, we could illustrate {\textit{how automatically designed structures would compare to previous manually designed EEG-based deep neural networks}}. 
The results are shown in Table~\ref{tb:mix_results}, where $^{*}$ denotes the results from our reproduction.
\begin{table}[hbpt]
\centering

\caption{\label{tb:mix_results} {Mean accuracy on BCI Competition IV-2a dataset, where $^{*}$ denotes the results are from our reproduction. \textbf{{CTDARTS-\tiny{EEG}$^{\star}$}} denotes our proposal without network scale constraint.}} 
\setlength{\tabcolsep}{0.7mm}{}
\begin{tabular}{llll}
\hline
Methods            & Accuracy. \tiny{(\%)} & Parameter \tiny{Scale}.  & MACs \\ \hline
EEGNet$^{*}$            & $67\sim72$      & $2.64k$    & $12.9M$           \\
Shallow-Conv$^{*}$    & $74\sim75$        & $47.3k$   & $63.5M$  \\
FBCSP              & $73.70$                & $261k$    & $104M$   \\
Riemanian$^{*}$    & $72\sim74$          & $50.0k$   & $-$       \\
MSFBCNN            & $75.80$                & $155k$    & $202M$       \\
EEG-TCNET          & $77.34$                & $20.5k$   & $6.8M$          \\
TPCT               & $88.7$                 & $7.78M$   & $1.73G$       \\
\hline
\textbf{CTDARTS\tiny{-EEG}}            & \small{$76.8\sim79.1$}       & \small{$18.2\sim32.1k$}   & \small{$32\sim280M$} \\
\textbf{CTDARTS\tiny{-EEG}$^{\star}$}    & \small{$77.3\sim80.4$}       & \small{$24.1\sim120k$}   & \small{$28\sim107M$}\\
\hline
\end{tabular}
\end{table}

\begin{table*}[hbpt]
\caption{\label{tb:withinsubjectp}Subject dependent performance on BCI Competition IV-2a datasets.}
\centering
\setlength{\tabcolsep}{1.5mm}{}
\small
\begin{tabular}{lllllllllllllll}
\hline
     & \multicolumn{6}{c}{Unified Weights}                                                                                                                                                               & \multicolumn{6}{c}{Subject-Specific Weights}                                                                                                                                                      & \multicolumn{2}{c}{Variable Structure}                 \\ \cline{2-15} 
     & \multicolumn{2}{c}{EEGNet$^*$}                                 & \multicolumn{2}{c}{EEG-TCNet}                                  & \multicolumn{2}{c|}{CTNAS-EEG}                                      & \multicolumn{2}{c}{EEGNet$^*$}                                 & \multicolumn{2}{c}{EEG-TCNet}                                  & \multicolumn{2}{c|}{CTNAS-EEG}                                      & \multicolumn{2}{c}{CTNAS-EEG}                                      \\
     & \multicolumn{1}{c}{Acc.\tiny{$^{(\%)}$}} & \multicolumn{1}{c}{$\kappa$} & \multicolumn{1}{c}{Acc.\tiny{$^{(\%)}$}} & \multicolumn{1}{c}{$\kappa$} & \multicolumn{1}{c}{Acc.\tiny{$^{(\%)}$}} & \multicolumn{1}{c|}{$\kappa$} & \multicolumn{1}{c}{Acc.\tiny{$^{(\%)}$}} & \multicolumn{1}{c}{$\kappa$} & \multicolumn{1}{c}{Acc.\tiny{$^{(\%)}$}} & \multicolumn{1}{c}{$\kappa$} & \multicolumn{1}{c}{Acc.\tiny{$^{(\%)}$}} & \multicolumn{1}{c|}{$\kappa$} & \multicolumn{1}{c}{Acc.\tiny{{$^{(\%)}$}}} & \multicolumn{1}{c}{$\kappa$} \\ \cline{2-15} 
S1   & 85.12                           & 0.74                         & 85.77                           & 0.81                         & 87.35                           & \multicolumn{1}{l|}{0.78}     & 87.09                           & 0.82                         & 89.32                           & 0.86                         & 93.25                           & \multicolumn{1}{l|}{0.80}     & 95.67                           & 0.87                         \\
S2   & 61.05                           & 0.32                         & 65.02                           & 0.53                         & 72.41                           & \multicolumn{1}{l|}{0.58}     & 62.38                           & 0.52                         & 72.44                           & 0.63                         & 80.52                           & \multicolumn{1}{l|}{0.71}     & 81.35                           & 0.76                         \\
S3   & 86.36                           & 0.79                         & 94.51                           & 0.93                         & 90.23                           & \multicolumn{1}{l|}{0.81}     & 93.21                           & 0.91                         & 97.44                           & 0.97                         & 98.15                           & \multicolumn{1}{l|}{0.92}     & 98.01                           & 0.92                          \\
S4   & 65.77                           & 0.67                         & 64.91                           & 0.53                         & 67.22                           & \multicolumn{1}{l|}{0.58}     & 74.57                           & 0.58                         & 75.87                           & 0.68                         & 79.65                           & \multicolumn{1}{l|}{0.69}     & 81.20                           & 0.67                         \\
S5   & 66.90                           & 0.59                         & 75.36                           & 0.67                         & 76.10                            & \multicolumn{1}{l|}{0.71}     & 77.12                           & 0.70                         & 83.69                           & 0.78                         & 85.68                           & \multicolumn{1}{l|}{0.81}     & 86.55                           & 0.83                         \\
S6   & 54.01                           & 0.41                         & 61.40                           & 0.49                         & 62.60                           & \multicolumn{1}{l|}{0.44}     & 62.10                           & 0.51                         & 70.69                           & 0.61                         & 76.24                           & \multicolumn{1}{l|}{0.60}     & 77.35                           & 0.69                         \\
S7   & 87.93                           & 0.88                         & 87.36                           & 0.83                         & 89.13                           & \multicolumn{1}{l|}{0.84}     & 90.25                           & 0.89                         & 93.14                           & 0.91                         & 93.89                           & \multicolumn{1}{l|}{0.89}     & 94.56                           & 0.90                          \\
S8   & 77.34                           & 0.81                         & 83.76                           & 0.78                         & 79.21                           & \multicolumn{1}{l|}{0.76}     & 86.57                           & 0.85                         & 86.71                           & 0.82                         & 89.27                           & \multicolumn{1}{l|}{0.84}     & 90.05                           & 0.84                         \\
S9   & 75.21                           & 0.65                         & 78.03                           & 0.71                         & 77.68                           & \multicolumn{1}{l|}{0.66}     & 83.21                           & 0.70                         & 85.23                           & 0.80                         & 90.06                           & \multicolumn{1}{l|}{0.82}     & 91.88                           & 0.82                          \\ \cline{2-15} 
Avg. & 73.29                           & 0.651                        & 77.36                           & 0.698                        & 78.10                           & \multicolumn{1}{l|}{0.684}    & 79.72                           & 0.72                         & 83.84                           & 0.784                        & 87.41                           & \multicolumn{1}{l|}{0.786}    & 88.52                           & 0.809                        \\
Std. & 12.06                           & 0.186                        & 11.59                           & 0.155                        & 10.02                           & \multicolumn{1}{l|}{0.131}    & 11.50                           & 0.156                        & 9.20                            & 0.125                        & 7.39                            & \multicolumn{1}{l|}{0.102}    & 7.28                            & 0.088                        \\ \hline
\end{tabular}
\end{table*}

For the reproduced models, we run repeat experiments and report the upper and lower bound of each model.
Since for CTNAS-EEG, the searched architecture is different with each searching process, we directly report the range of the parameter scales in Table~\ref{tb:mix_results}. 
It could be observed that the search algorithm could reach a performance range of $77.0\sim79.1\%$ which could outperform the current SOTA light-weighted model EEG-TCNET with an accruacy $77.34\%$ while keeping the parameter scales comparable ($18.2k\sim32.1k$ vs. $20.5k$). 
The best performance of our proposal with or without structure constraint could respectively reach $79.1\%$ and $80.4\%$ on average accuracy on the mixed dataset while keeping the structure scale at $24.2k$ and $61.6k$.

\begin{table}[t]
\centering
\setlength{\tabcolsep}{1mm}{}
\caption{\label{tb:emotionsd}Subject dependent performance on SEED Dataset, where \textbf{CTDARTS\tiny{-EEG}$^\star$} denotes our method without network scale constraint. Most previous works listed are based on DE features, however, our proposal is based on raw signal. }
\begin{tabular}{llll}
\hline
            & \multicolumn{2}{c}{Acc/Std. (\%)} \\ \hline
Methods     & SEED IV         & SEED V          \\ \hline
SVM\scriptsize{~\cite{suykens1999least}}       & 56.61/20.05     & 45.35/19.10     \\
RF\scriptsize{~\cite{breiman2001random}}        & 50.97/16.22     & 43.22/17.63     \\
CCA~\scriptsize{\cite{everitt2021encyclopedia}}         & 54.47/18.48     & 46.41/18.88     \\
GSCCA~\scriptsize{\cite{zheng2016multichannel}}      & 69.08/16.66     & 57.21/20.72     \\
DBN~\scriptsize{\cite{zheng2015investigating}}        & 66.77/07.38     & 56.82/17.23     \\
GRSLR~\scriptsize{\cite{li2019eeg} }     & 69.32/19.57     & 62.32/18.97     \\
GCNN~\scriptsize{\cite{defferrard2016convolutional} }        & 68.34/15.42     & 63.12/19.76     \\
DGCNN~\scriptsize{\cite{song2018eeg}}       & 69.88/16.29     & 65.34/20.36     \\
BiDANN~\scriptsize{\cite{li2018novel}}      & 70.29/12.63     & -               \\
A-LSTM~\scriptsize{\cite{song2019mped}}      & 69.50/15.65     & 65.10/20.20      \\
BiHDM~\scriptsize{\cite{li2020novel}}       & 74.35/14.09     & 72.33/17.65     \\
DEFuzzy~\scriptsize{\cite{zhang2020hierarchical}}     & 70.34/16.73     & 70.08/18.21     \\ \hline
\multicolumn{1}{l}{\multirow{2}{*}{\begin{tabular}[c]{@{}l|@{}}Specific\\ Weights\end{tabular}} \multirow{2}{*}{\begin{tabular}[c]{@{}l@{}} \textbf{CTDARTS\tiny{-EEG}}\\ \textbf{CTDARTS\tiny{-EEG}$^\star$} \end{tabular}}} & 75.89/15.65     & 73.04/17.74     \\
\multicolumn{1}{l}{}   & 76.16/17.42     & 74.26/20.08     \\
 
 \hline
\multicolumn{1}{l}{\multirow{2}{*}{\begin{tabular}[c]{@{}l|@{}}Variable\\ Structure\end{tabular}} \multirow{2}{*}{\begin{tabular}[c]{@{}l@{}} \textbf{CTDARTS\tiny{-EEG}}\\ \textbf{CTDARTS\tiny{-EEG}$^\star$} \end{tabular}}} & 76.21/15.96    & 73.89/18.29     \\
\multicolumn{1}{l}{} & 77.01/17.29     & 74.21/19.43     \\ \hline
\end{tabular}
\end{table}

\subsubsection{Subject Dependent Performance}
\label{subsec:dependentsubjects}

One unique advantage of the proposed network searching method is that CTNAS-EEG could customize the network not only for different tasks but also for different subjects. 
Under this setting, the model could access the subject information during training, and fit the network with each human subject specifically, which is the first framework that could realize customization for each human user. 
We compare the proposed CTNAS-EEG with previous baselines with three evaluation methods as below and report the results in Table~\ref{tb:withinsubjectp}.

\begin{itemize}
    \item \textbf{Unified Weights} suggest that we evaluate by fixing one certain structure and training unified weights for all subjects. In {Unified Weights} part, where EEGNet, EEG-TCNet, and CTNAS-EEG respectively reach $73.29\%$, $77.36\%$, and $78.10\%$ average accuracy on four-class motor imaginary classification. 
    The results indicate the proposal has a slight advantage over previous baseline approaches, yet the difference compared is small. Like previous observations~\cite{EEGNet_Fusion}, the accuracy can vary severely on different subjects, such as on S2, S4, and S5, all three methods achieve significantly lower accuracy compared to other subjects.
    The performance gap still suggests that a unified model might not be generally good for all subjects. 
    \item \textbf{Subject-Specified Weights}\footnote{Previous methods could improve within-subjects performance by separately training the same model structure for each subject. We strictly follow the previous setting for a fair comparison.} fix one certain structure and search for specific weights.
    The results are shown as Subject-Specific Weights part in Table~\ref{tb:withinsubjectp}, where EEGNet, EEG-TCNet and CTNAS-EEG respectively reach $79.72\% (+6.43\%)$, $83.84\% (+6.48\%)$ and $87.41\% (+9.31\%)$ average accuracy. All three methods have significant performance improvement after subject-specific training.
However, the proposed CTNAS-EEG has the highest performance increase ($+9.31\%$ vs. $+6.43\%$ and $+6.48\%$). 
Meanwhile, the accuracy standard deviation of CTNAS-EEG on subjects is $7.39\%$, which is also significantly lower than baseline methods ($11.50\%$ and $9.20$), which suggests that CTNAS-EEG has a smaller performance gap between different subjects. 
It suggests that our searched structure could provide better optimization for previously ‘harder subjects’.
    \item \textbf{Variable Structure} denotes customizing specific structure $\&$ weights for each subject, which is enabled by the unique properties of our proposal. It could realize more detailed optimization for specific structures. 
    It is observed that, by customizing the structure for each subject, the average accuracy and standard deviation respectively reach $88.52\%$ and $7.28\%$. 
    It suggests that customizing the structure for each structure could slightly improve model performance, yet the impact is not dominated. 
    This is rational since all these subjects' samples are motor-imaginary data, it should not require extremely different operators or structures while only changing the subject under the same task.
    However, CTNAS-EEG provides the ability to customize the structure for each subject and reach SOTA performance for follow-up researchers' reference. 
\end{itemize}

\subsection{Emotion Performance}
\label{subsec:performance Emtion}

As the emotion tasks have severe subject differences naturally, previous works~\cite{li2020novel} are mostly evaluating results based on subject-dependent performance. 
Thus, we directly evaluate subject-depended performance on Emotion tasks.
The experimental results are reported in Table~\ref{tb:emotionsd}.
As SEED datasets have more subjects than BCI Competition IV datasets, we directly report the average and standard deviation metrics in Table~\ref{tb:emotionsd}.  
Since this work focuses on the structure, we didn't bring multi-modality input to increase model performance, where all results reported is based on raw EEG signal input. 
The CTNAS-EEG could respectively reach $76.21\%$, $77.01\%$ on SEED-IV dataset, and reach $73.89\%$, $74.21\%$ on SEED-V dataset with or without network scale constraint. 
This result supports that the proposed CTNAS-EEG could unify separated MI and Emotion tasks within one simple framework and both reach SOTA performance.
Thus, we could first observe and compare the search operators cross-task wise, where the analysis is reported and analyzed in Section~\ref{subsec:crosstask_op}.

\subsection{Cross-Task Searching Analysis}

\label{subsec:crosstask_op}

As mentioned in Section~\ref{subsec: Searchspace} that searching is learning to select proper operators\footnote{Please denote that `none' operator denotes no connection between a node pair, which may change the network topology.} from the search space. 
Unifying searching cross tasks, we can compare suitable operators and their difference between MI and Emotion. 
The difference is visualized in two aspects, 1) statistical results of selecting all operators and 2) the probability distribution variance. 


\subsubsection{Statistical results of all operators}
\label{subsubsec:stat_operators}
\begin{figure}[t]{
\centering
\subfigure[Results on MI datasets]{
\includegraphics[width=0.47\textwidth,height=3.5cm]{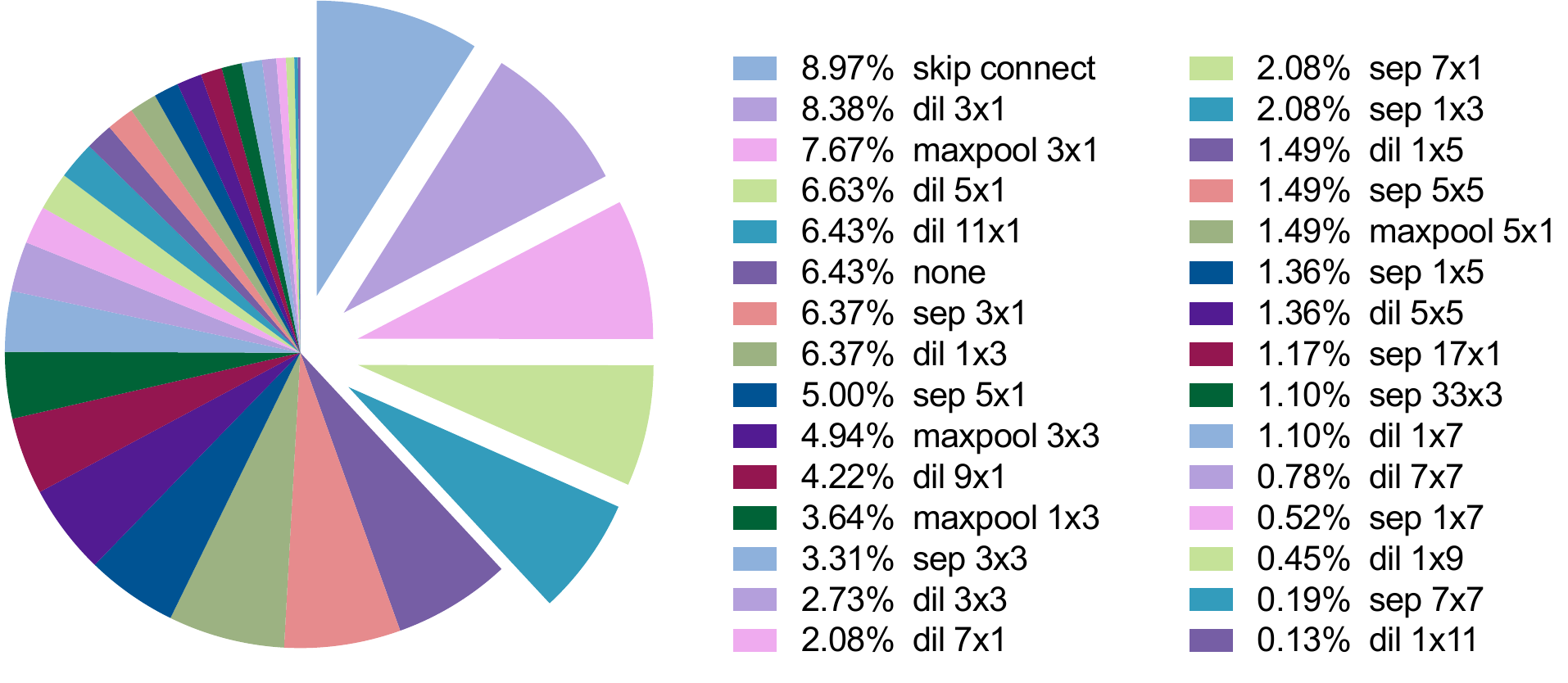}\label{fig:mixed_vis}
}
\subfigure[Results on Emotion datasets]{
\includegraphics[width=0.47\textwidth,height=3.5cm]{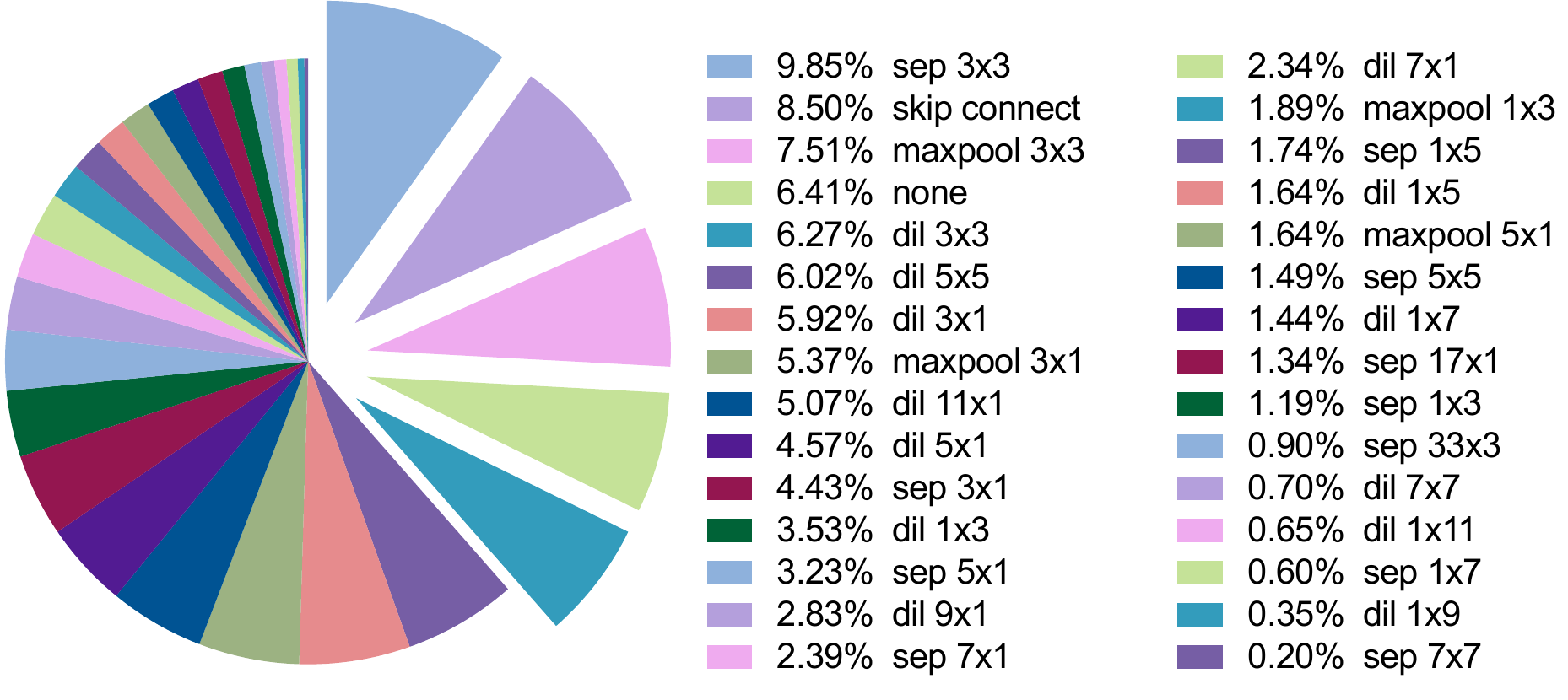}\label{fig:ss_vis}
}
\caption{\label{fig:stat} The statistical results of operator selected out from the search space, where `dil', `sep' respectively denotes dilation convolution and separable convolution. }
}
\end{figure}

To analyze how each operator is useful for different tasks, we directly count the numbers of each operator selected as part of the validation structure~\footnote{Means searched structures with performance no lower than our best results by $10\%$.} after each epoch of the training.  
The statistical includes all trails of unified searching and subject-specific searching reported in Section~\ref{subsec:performance MI} and Section~\ref{subsec:performance Emtion}. 
For the statistical analysis, we include all possible operators in the large search space, where the results are shown in Figure~\ref{fig:stat}.  
For MI tasks, `narrow' operators are more likely to be selected, such as dilation convolution `dil 3x1', `dil 5x1', `dil 11x1', `sep 3x1', and `maxpool 3x1', which suggest that operators with time-wise aggregation have a higher probability to be selected from searching. 
It could also be observed that the dilation convolution shows advantages over separate convolution and normal convolution. 
We argue this phenomenon is rational since the dilation operators could provide a larger perception field, which is important for extracting long-term time-series MI signals.
For Emotion tasks, square operators such as `sep 3x3' and `maxpool 3x3', take advantage of others. 
It suggests that Emotion tasks may need more abilities to extract information across different time slices. 
It is also rational since the emotion task is not that `clear' compared to MI tasks, and may need more samples from a different time domain to extract confident features.



\begin{figure}
    \centering
    \includegraphics[width=0.5\textwidth,height=9.5cm]{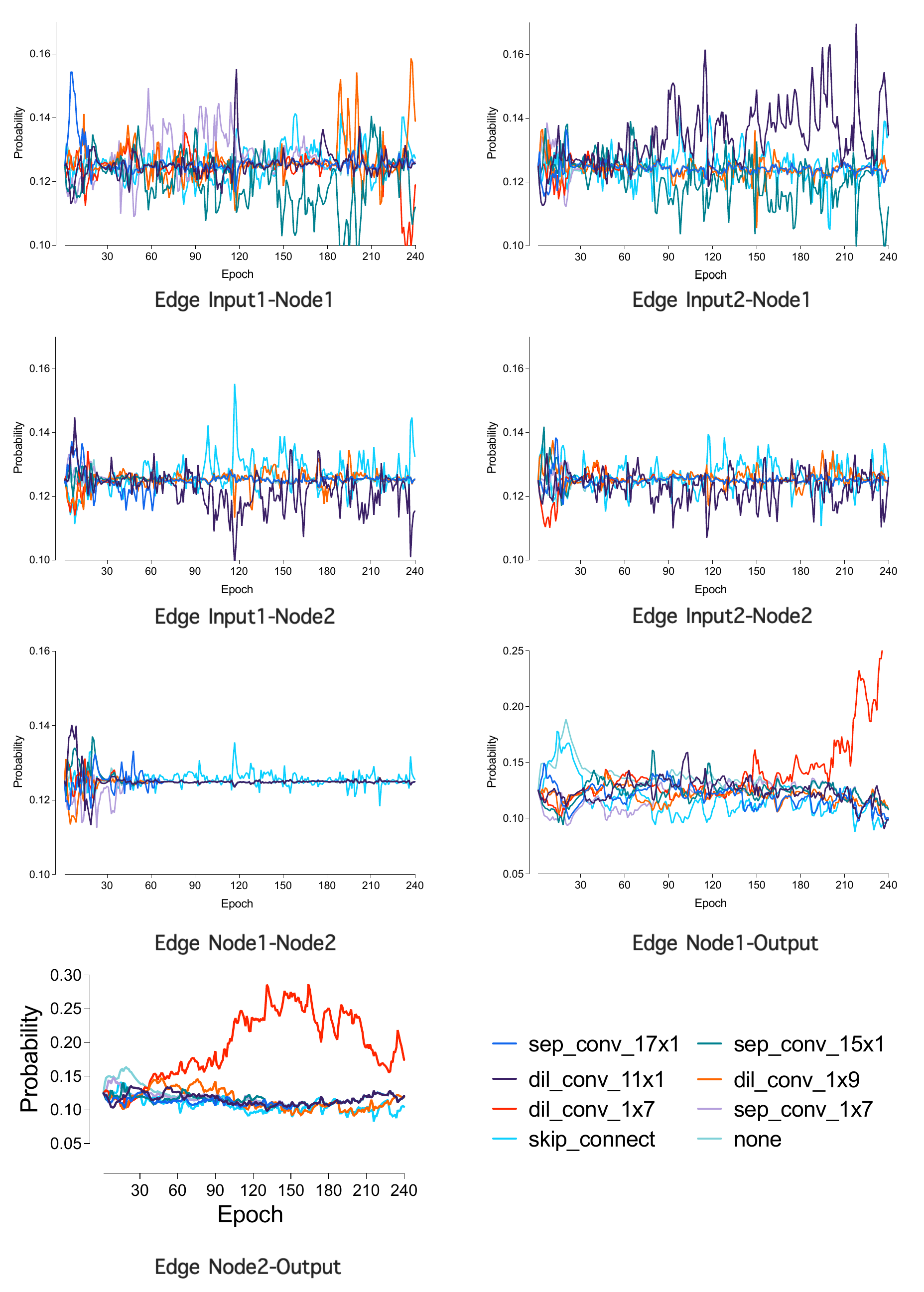}
    \caption{Probability distribution of operators through training on BCI Competition IV dataset from specific subject (S3). Input$\{1,2\}$, Node$\{1,2\}$, and output denote different parts described in Figure~\ref{fig:metanetwith_searchspace}}
    \label{fig:customizeddoperatordist_sp}
\end{figure}

\subsubsection{Operator variance considering topology}
\label{subsubsec:Operator variance trhough training}
Since the operator between different node pairs may require different properties, we visualize the operators' distribution variance on each edge of the search-able cell.
It should be noted that we only listed the top 8 operators in our search space for a clear presentation.
We keep the hyper-parameters as well as nodes in each search-able cell the same as the performance evaluation experiments.
Here we selected subject 3 (S3) in MI as representative of subject customized structure search as S3 is one of the best subjects in Table~\ref{tb:withinsubjectp}. 
Here, the `none' operator denotes that there is no actual calculation of our feature flow between the given node pair. 
Here we selected subject 3 (S3) in the BCI Competition IV dataset to visualize the operators' distribution change of subject customized structure search as shown in Figure~\ref{fig:customizeddoperatordist_sp}. 
It is observed that, at `shallow' layers, the degree of separation between operators is not that good, which suggest that various operators are suitable to extract shallower features from raw EEG signal. 
However, on deeper layers, the operators' distribution is more separable, such as `dil 1x9' stands out to aggregate information from different time slices. 
Limited to paper length, we provide more visualization results in Appendix Section~\ref{ap:visearchp}. 


\section{Conclusion}

This paper proposes the CTNAS-EEG framework to adopt neural architecture search into EEG signal processing. 
CTNAS-EEG is the first approach that unifies multiple separated tasks and could customize model structure for each human subject.
Experimental results suggest our proposal could reach SOTA performance on both MI and Emotion tasks, which could be convenient for researchers unfamiliar with specific network design in the BCI area. 
Besides, cross-task searching analysis is provided to give follow-up researchers as good references. 


\begin{quote}
\begin{small}
\bibliography{aaai23}
\end{small}
\end{quote}

\clearpage
\appendix

\twocolumn[
\begin{@twocolumnfalse}
\section*{\centering{Supplementary Material: 1599
Cross Task Neural Architecture Search\\ for EEG Signal Recognition}}

\section*{\Large Paper ID: 1599 }

\vspace{20pt}

\end{@twocolumnfalse}
]

\vspace{25pt}


Appendix~\ref{subsec: Preliminaries} provides preliminary knowledge of differentiable architecture search.
Appendix~\ref{ap:searchoperators} provides the detailed search space we use for cross-task EEG classification.
Appendix~\ref{ap:implementation} presents more detailed experimental settings of CTNAS. Appendix~\ref{ap:visearchp} presents 1) visualization of the operators' change during the search period 2) visualization of the final searched structures for different tasks. 
Appendix~\ref{subsec:hyperparam} provides an ablation study of hyper-parameters in our framework, including search space scale, batch size, and the effect of network constraints. 

\vspace{20pt}

\section{Preliminaries}
\label{subsec: Preliminaries}

The key factor of constructing neural architecture search (NAS) contains three aspects, search space, search strategy (optimization strategy), and evaluation metric. 
These three aspects respectively indicate the candidates and compilation of potential structures for searching, how we search and update the structures in defined search space, and how we decide which structure to choose during the searching process. 
Previous explorations mostly in the image area have revealed that if the search space is too large, the searching process would lead to severe time consumption and hardness of convergence. 
A modern formation is simply to connect various blocks and only search operators and their connection inside one certain block. \begin{figure}[!hbt]{
\centering
\subfigure[Mixed OP]{
\includegraphics[width=0.125\textwidth]{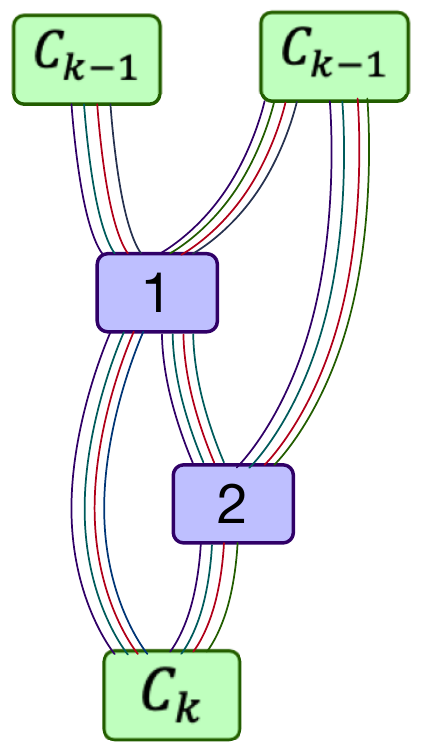}\label{fig:Mixed OP}
}
\subfigure[Argmax OP]{
\includegraphics[width=0.125\textwidth]{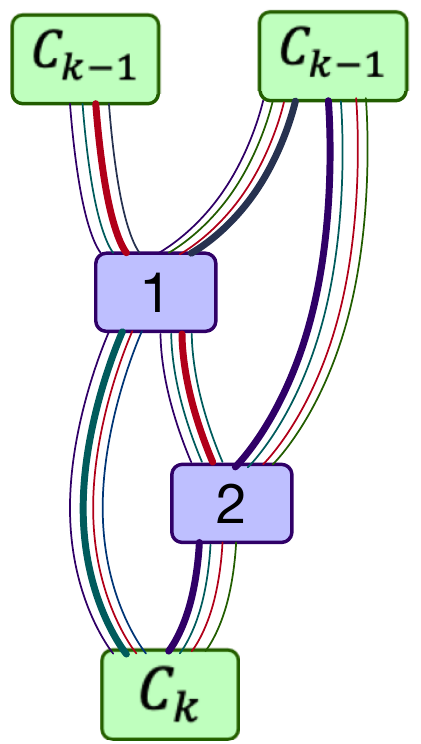}\label{fig:Argmax OP}
}
\subfigure[Structure]{
\includegraphics[width=0.125\textwidth]{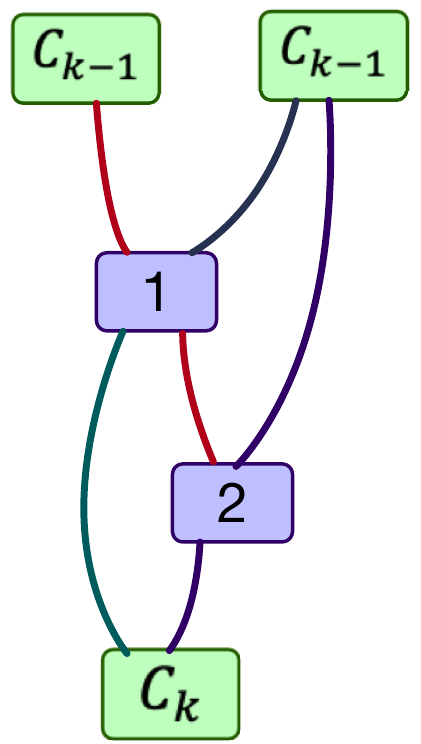}\label{fig:Complie Structure}
}
\caption{A schematic diagram of a differentiable architecture search cell, where the network cell is initialized with (a) mixed operators of search space between each node pair. Then the architecture is relaxed to continuous and optimized to selected (b) Argmax operator from the search space. (c) The structure is selected by compiling the operator with the maximized SoftMax probability. \label{fig:pre_darts}
}}
\end{figure}

In this work, similar to previous DARTS methods~\cite{DBLP:journals/corr/abs-1806-09055,DBLP:conf/eccv/ChuZZL20,DBLP:journals/corr/abs-1907-05737}, we use the search space as the building block as shown in Figure~\ref{fig:pre_darts}. 
The block could be represented as a directed acyclic graph \textbf{(DAG)}:
\begin{equation}
    G=(E,V), x_{i}\in V, i=1,2,...,N,c_{k-2},c_{k-1},c_{k}
\end{equation}
where N is the number of intermediate nodes inside each block. The block is defined to receive two input and one output we respectively set two input nodes as $x_{c_{k-2}}$, $x_{c_{k-1}}$, and one output node as $x_{c_{k}}$. 
Each node $x_i$ is a latent representation extracted from its input. Each edge associates with a operator $o_{(i,j)}$, where $o_{(i,j)}$ transforms $x_i$.
The feature transformation between node $x_i$ and $x_j$ could be defined as:
\begin{equation}
    x_j = \sum_{i<j} o_{(i,j)}(x_i)
\end{equation}
Here, the searching task could be defined as seeking the operator $o_{(i,j)}$ for each edge from search space $O$. The selection of one specific operator normally~\cite{DBLP:journals/corr/abs-1806-09055} is relaxed as SoftMax probability over all possible operators:
\begin{equation}\label{eq:mixedop}
      o_{(i,j)}(x) = \sum_{o^{'}\in O}\frac{{\rm exp}(\theta_{o}^{(i,j)})}{ \sum_{o^{'} \in O} {\rm exp}( \theta_{o^{'}}^{(i,j)})} o(x),
\end{equation}
where the weights of operators assigned for edge $(i,j)$ are defined as $\theta_{o}^{(i,j)}$, a probability vector of dimension $|O|$, the number of all operators. In that case, the structure searching problem could be simplified to optimize set of architecture parameters $\theta = {\theta_{o}^{(i,j)}}$. 
While the optimized architecture parameters $\theta$ are finalized through searching, the network structure could be obtained by replacing each mixed operation to operators with the highest probability ${\rm argmax}_{o \in O} \theta_{o}^{(i,j)}$.
Under this setting, the goal is jointly to optimize architecture parameters $\theta$ and the network weights $w$. 
However, due to the complexity of strict joint optimization problems, most works relax this problem to a bi-level optimization problem:
\begin{align}\label{eq:orgproblem} 
    &\mathop{\rm min}\limits_{\theta} \mathcal{L}_{\rm val}(w^{*}(\theta),\theta) \\\nonumber
    s.t. \quad &w^{*}(\theta)={\rm arg min}_{w} \mathcal{L}_{\rm train}(w(\theta)),
\end{align} 

where the object is to find optimal structure $\theta$ for minimal loss function on the validation set $\mathcal{L}_{\rm val}$ while given local optimal $w^{*}$ by fixing local optimal $\theta$ from previous optimization step by minimizing loss function $\mathcal{L}_{\rm train}$ on the training set. 
Since this work focuses more on introducing neural architecture search methods into EEG signal processing, here we do not modify the basic problem setting of DARTS~\cite{DBLP:journals/corr/abs-1806-09055}.
Essentially, we use this setting to make the architecture itself a set of ’hyper parameters’, thus making it searchable. 
The motivation of this work is to introduce automatic searching methods to overcome compared less human empirical design exploration in EEG signal processing.

\section{Detailed Operators in Search Space}
\label{ap:searchoperators}

As it has mentioned in Section~\ref{subsec: Preliminaries} and Equation~\ref{eq:mixedop}, during the searching period, features between each node pair are weighted by mixed operation. 
In that case, efficient and compatible are important at the same time.
If the search space contains many inefficient operators, the `bad' operators would bring negative effects to the descending of the MetaNet.
Meanwhile, if a search space is not compatible for multiple tasks, the searching algorithm can't select proper operators as it is not existing in the search space. 
Especially, considering the proposed CTNAS-EEG is supposed to perform cross-task searching, a larger search space is more important. 
EEG signals contain information on both time-series and spatial aspects.
A typical input formation of the raw input signals collected via the 10-20 system has shape $bs*{\rm datapoints}*{\rm channels}$, where data points contain more time-series electrode variances while channels denote spatial information of different electrodes on the human brain.
Given the sampling rate of 250hz, a typical sample may contain information within couple of seconds.
To avoid the input signal becomes too 'long', we slice the data points dimension of the whole matrix into fragments with intersections and stack these fragments in another dimension, after which, the input signal is processed from $bs*{\rm datapoints}*{\rm channels}$ to $bs*{\rm slices}*{\rm datapoints}*{\rm channels}$.
Considering the computational efficiency, we mostly use convolution operators to design our search space.

\begin{table}[hbpt]
\caption{Detailed search space of CTNAS-EEG\label{tb:searchspace}}
\begin{tabular}{|l|l|}
\hline
\textit{\textbf{Type} }                                                           &\textit{\textbf{ Operators    }}                                                                                                                                                                      \\ \hline
Pooling                                                         & \begin{tabular}[c]{@{}l@{}}maxpool 3x3,  maxpool 3x1, \\ maxpool 5x1,  maxpool 1x3\end{tabular}                                                                                    \\ \hline
\begin{tabular}[c]{@{}l@{}}Seperable\\ Convolution\end{tabular} & \begin{tabular}[c]{@{}l@{}}sep 3x1, sep 5x1, sep 7x1, sep 11x1, \\ sep 17x1, sep 3x3, sep 5x5, sep 7x7, \\ sep 33x3, sep 17x3,sep 1x3, sep 1x5, \\ sep 1x7, sep 1x11\end{tabular}  \\\hline
\begin{tabular}[c]{@{}l@{}}Dilation\\ Convolution\end{tabular}  & \begin{tabular}[c]{@{}l@{}}dil 3x1, dil 5x1, dil 7x1, dil 11x1, \\ dil 17x1, dil 3x3, dil 5x5, dil 7x7, \\ dil 17x3, dil 33x3, dil 1x3, dil 1x5, \\ dil 1x7, dil 1x11\end{tabular} \\ \hline
Others                                                          & none, skip connection                                                                                                                                                              \\ \hline
\end{tabular}
\end{table}

To better adapt normal convolutional kernels to time-series signals, we provide a series of ‘narrow’ and ‘Flat’ kernels.
For the `Narrow' kernels, we provide various lengths of kernels $\{3,5,7,11,17\}\times1$ respectively corresponding to different time-series perception fields. Considering the 250Hz sampling rate, the search space could cover a time window between 12ms to 108ms. Considering the feature flow in the whole network structures, the perception field could be extended to the whole period of the reaction time which is 1.5 seconds. 
For `Flat' kernel, we provide various lengths of kernels $1\times\{3,5,7,9,11\}$ corresponding to different EEG channels. 
The detailed search space we used is listed in Table~\ref{tb:searchspace}


\section{Implementation Details}
\label{ap:implementation}
Since the training samples of the BCI IV datasets are limited, we used compared slim search space for training, where the numbers of blocks, and nodes in each blocks defined in Section~\ref{subsec: Searchspace} are respectively set to 3 and 3. 
Since the singal-to-norise-rate (SNR) of EEG singals is low, we used compared slim search space for training, where the numbers of blocks, and nodes in each blocks defined in Section~\ref{subsec: Searchspace} are respectively set to 3 and 2. 
We use Adam~\cite{zhang2018improved}, and SGD~\cite{bottou2012stochastic} respectively for updating structure parameters and network weights. 
For the Adam optimizer, we use an initial learning rate of 0.01 with a multi-step learning rate decay strategy, where the beta value is between 0.5 to 0.99. For SDG optimizer, we use an initial learning rate of 0.01 with Cosine learning rate decay to a minimum value of 0.0001.
Normal dropout with dropout rate 0.1 is used in the training process. 
The structure parameters and the network weights are updated step by step on a 1:1 ratio. 
The whole search algorithm is implemented in PyTorch\footnote{\href{https://pytorch.org/}{https://pytorch.org/}}, a convenient deep learning library based on python.
We \textbf{release the code on Github to contribute to the open-source community}~\footnote{ \href{https://github.com/DuanYiqun/CTNAS-EEG}{https://github.com/DuanYiqun/CTNAS-EEG}}. 
Each operator in our search space is implemented in a simple block, where before the operation we put the Elu~\cite{clevert2015fast} activation function, and after the operation we normalize the output feature map by the batch norm.
The networks are trained on 4 Nvidia V100 GPUs, where in order to acquire larger batch size, we also apply distributed training with normal all-reducing gradient synchronization.

\section{Visualization of Searching Process}

\label{ap:visearchp}


\subsection{Operator variance considering topology}
The mentioned experimental results in main paper Section~\ref{sec:exp} includes both unified structure search and structure search customized for each subject.
In main results we only give an example of visualizing searching process on subject 3 (s3). 
However in Appendix, we further compare the visualization between unified structure search (on mixed subjects) and subject specific structure search. 
As the operator between different node pairs may require different properties, we first visualize the probability change on each edge of the search-able cell respectively on MI and Emotion under both unified search setting and subject-specific setting.

\begin{figure*}[!hbt]{
\centering
\subfigure[Unified search on mixed subjects.]{
\includegraphics[width=0.47\textwidth,height=9cm]{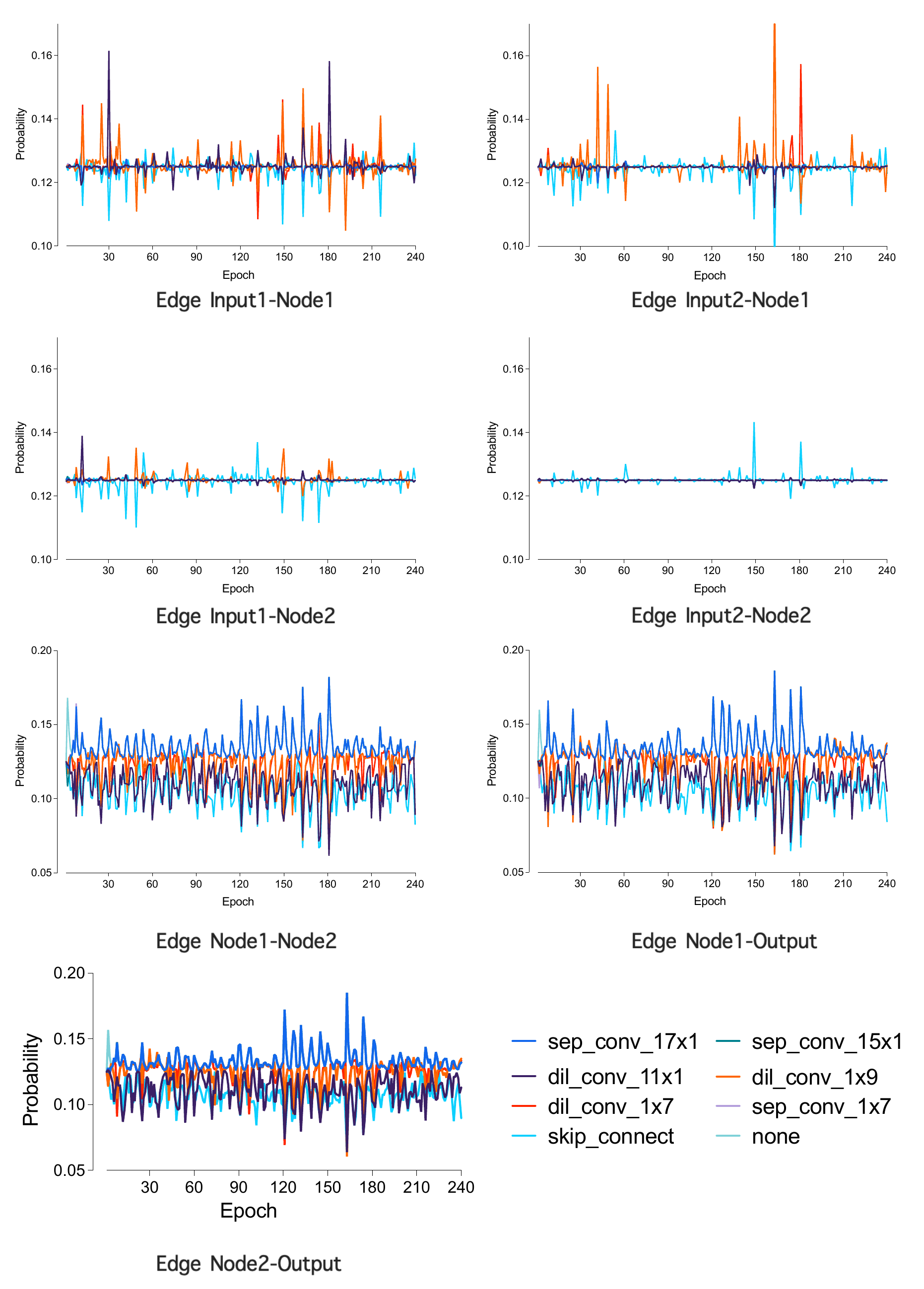}\label{fig:mi_mix_search}
}
\subfigure[Subject specific search on S3]{
\includegraphics[width=0.47\textwidth,height=9cm]{img/subject_2_operator_layout-eps-converted-to.pdf}\label{fig:mi_subject_search}
} 
\caption{Operator distribution change on BCI Competition IV dataset during searching considering topology, where Figure~\ref{fig:mi_mix_search} and \ref{fig:mi_subject_search} respectively denotes unified search and subject-specific search. \label{fig:visualizationonboth}
}}
\end{figure*}

\subsubsection{BCI Competition IV Dataset for MI}
\label{subsubsec:Operator variance trhough training MI}

It should be noted that we only listed the top 8 operators in our search space for a clear presentation.
We keep the hyper-parameters as well as nodes in each search-able cell the same as the performance evaluation experiments reported in Table~\ref{tb:mix_results} and Table~\ref{tb:withinsubjectp}.
Here, the `none' operator denotes that there is no actual calculation of our feature flow between the given node pair. 
Here we selected subject 3 (S3) as representative of subject customized structure search as S3 is one of the best subjects in Table~\ref{tb:withinsubjectp}. 
The results of the unified search and subject-specific search are respectively shown in Figure~\ref{fig:mi_mix_search} and Figure~\ref{fig:mi_subject_search} respectively.
It could be observed that, compared to a subject-specific search where certain operators would take clear advantages during training, the unified search leads to more evenly distributed operator probabilities.
Although operators `dil\_conv\_1$\times$7' and `sep\_conv\_15$\times$1' have a higher probability to be selected out, the distribution advantage is somehow `not definite', especially in `shallower'\footnote{Here, `shallower' and `deeper' respectively denotes the edge is located on beginning or ending of the cell, which is a directed acrylic graph.} edges. 
This phenomenon suggests two interesting findings. 
1) The subject-mixed datasets do have higher diversity so the searching algorithm is hard to choose several certain operators during the search period. 
2) Multiple operator choices could reach similar performance simultaneously, which suggests that the desired efficient structure is not unique, especially for those in `shallower' layers. 
Many similar structures could take the same effects, especially for unified search mixed subjects.

\begin{figure*}[t]{
\centering
\subfigure[Unified search on mixed subjects.]{
\includegraphics[width=0.48\textwidth,height=8.5cm]{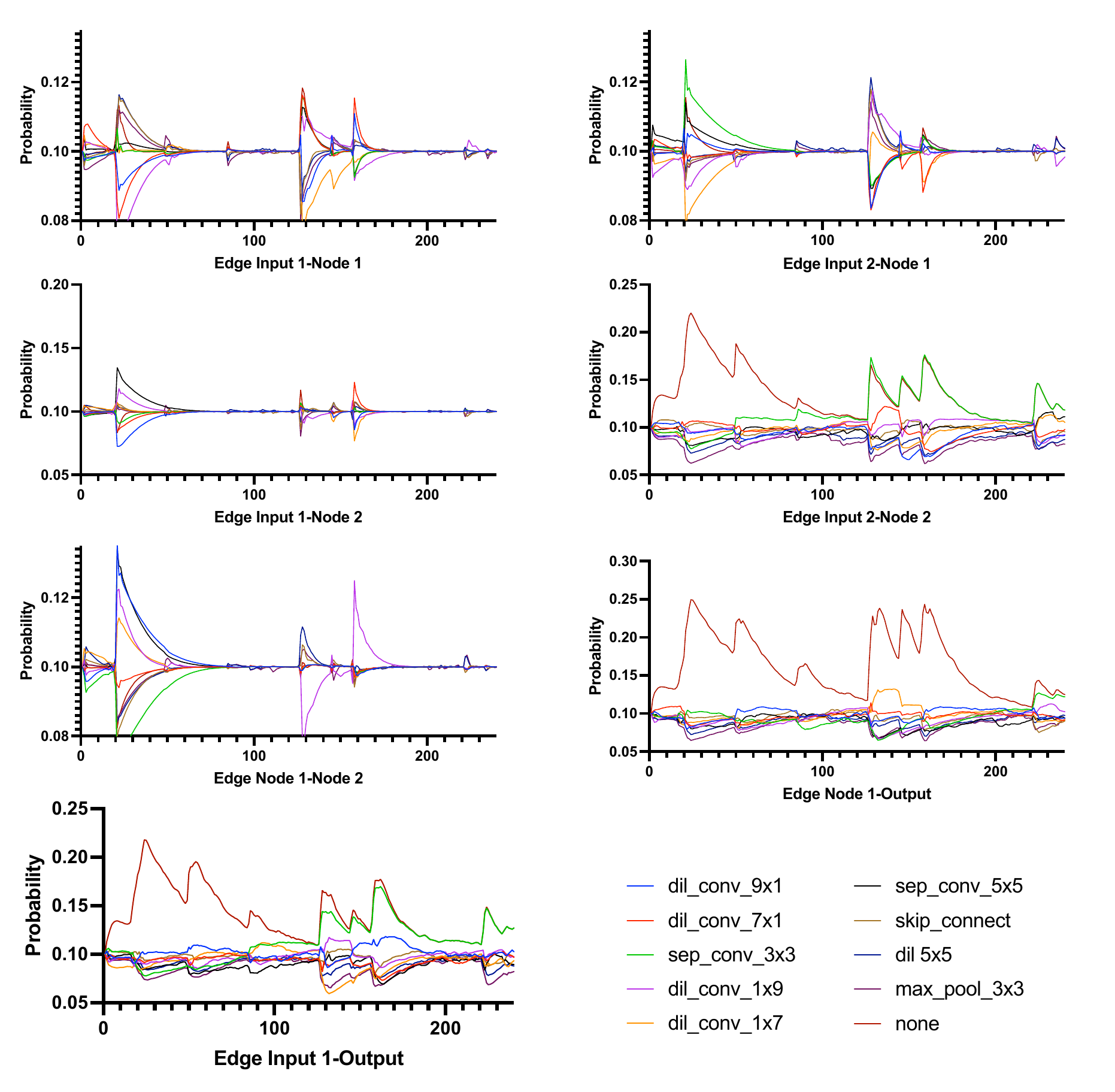}\label{fig:mi_mix_search_seed}
}
\subfigure[Subject specific search on S3]{
\includegraphics[width=0.48\textwidth,height=8.5cm]{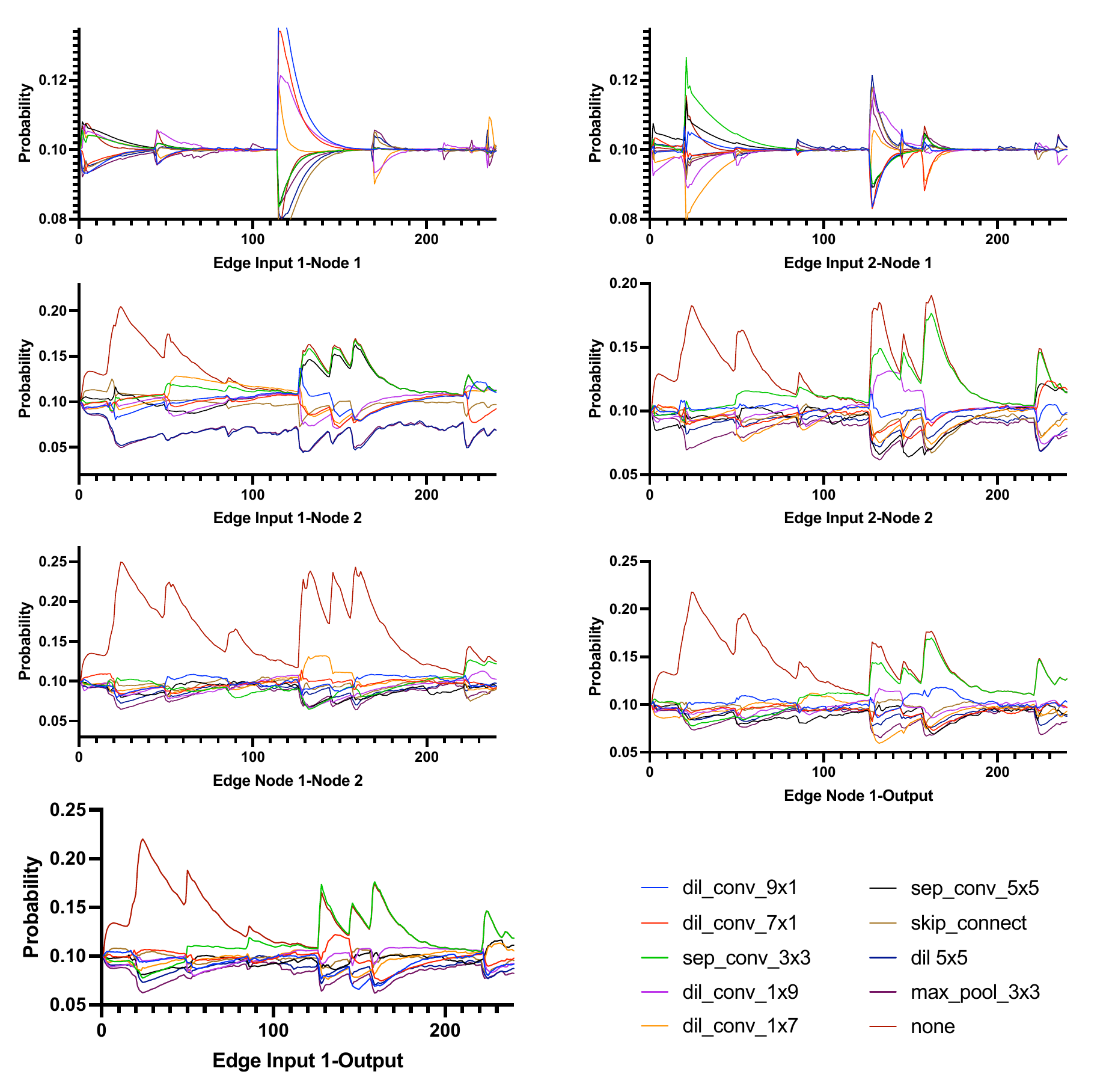}\label{fig:mi_subject_search_seed}
} 
\caption{Operator distribution change on SEED V dataset during searching considering topology, where Figure~\ref{fig:mi_mix_search_seed} and \ref{fig:mi_subject_search_seed} respectively denotes unified search and subject-specific search. \label{fig:visualizationonboth_seed}
}}
\end{figure*}

\subsubsection{SEED V Dataset for Emotion}
\label{subsubsec:Operator variance trhough training Emotion}
Similarly, we visualize the operator distribution change during searching on the Emotion task as well to compare with the MI task. 
We choose SEED V dataset to visualize the searching process, where both mixed-subjects searching (Figure~\ref{fig:mi_mix_search_seed}) and subject-specific searching (Figure~\ref{fig:mi_subject_search_seed}) is presented following basic setting mentioned above. 
It is observed that the operator distribution of MI and Emotion tasks has a clear difference. 
The operator distribution of MI is more constant and stable, while the probability of different operators is distributed more evenly. 
However, for Emotion tasks, the network forms operator preferences quickly but also with dramatic change, which is supported by the `pulse-like' curves reported in Figure~\ref{fig:visualizationonboth_seed}. 
This phenomenon is observed in both unified search and subject-specific search on the SEED V dataset. 
It suggests that Emotion tasks are more `picky' than the MI tasks, which means it's harder to select proper structures for Emotion tasks.
This observation is also supported by the final searched performance reported in the main paper Table~\ref{tb:withinsubjectp} and Table~\ref{tb:emotionsd}, which were respectively given the same 4 class~\footnote{BCI competition IV-2a for MI and SEED IV for emotion.} classification tasks, the performance of MI could reach $88.52\%$ on average accuracy while the Emotion could only reach $76.16\%$. 
It is rational since the Emotion of a human subject is normally more confusing to justify compared to Motor Imaginary. 
This may introduce potential label noise. 
Meanwhile, the SEED dataset used in this paper stimulates human emotion by playing movie clips with emotional intention in front of human subjects, which may introduce unrelated stimulation from the movie contents. 
Still, one observation is consistent cross-task-wise, where the subject-specific search has a better degree of separation of selecting operators than the unified search. 
In other words, the algorithm is more confident and clear in searching structure for a certain subject rather than searching for a unified structure for all subjects. 
This is rational as all of these tasks share long-existing subject difference problems in the EEG area. 
Yet, experimental results in this paper support that searching for a specific subject is feasible and could reach competitive performance. 

\subsection{Visualization of Searched Structures}

\begin{figure*}[hbpt]{
\centering
\subfigure[Visualization of Mixed Search]{
\includegraphics[width=0.82\textwidth]{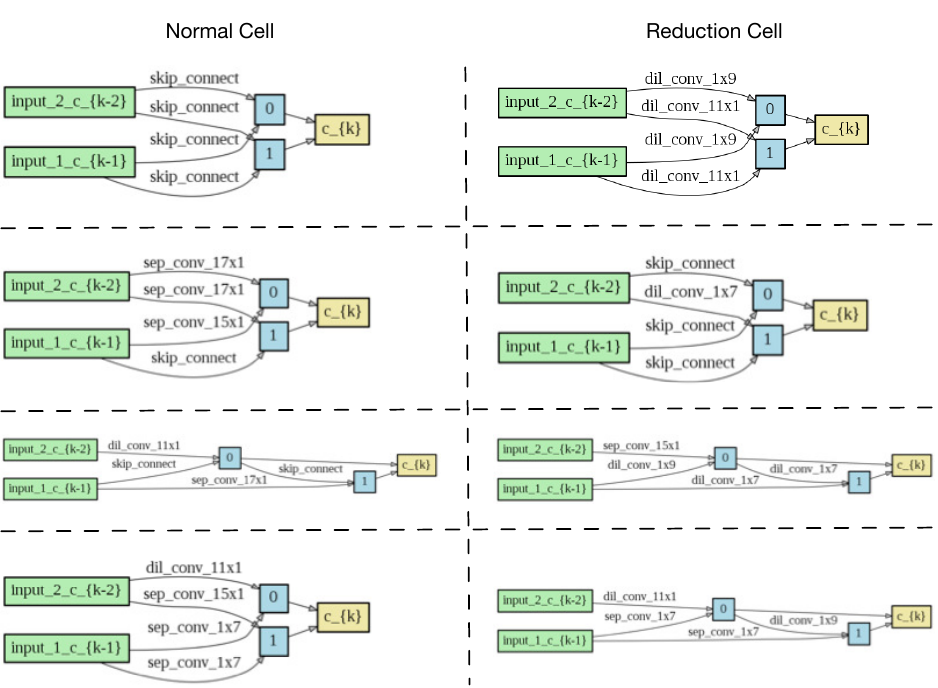}\label{fig:mixed_vis_bci4}
}
\subfigure[Visualization of Subject-Specific Search]{
\includegraphics[width=0.82\textwidth]{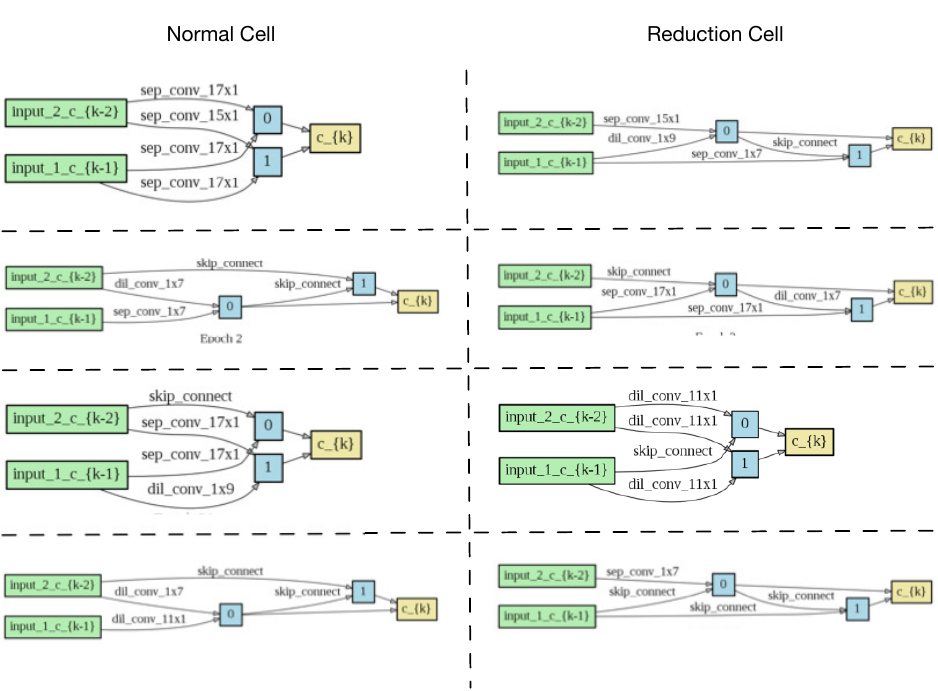}\label{fig:ss_vis_bci4}
}
\caption{\label{fig:vis} The searched structure on BCI Competition IV datasets, where Figure~\ref{fig:mixed_vis_bci4} and ~\ref{fig:ss_vis_bci4} respectively denotes the results on mixed-subject search and subject-specific search.}
}
\end{figure*}

\subsubsection{Searched Results on BCI Competition IV }
We visualize the searched whole network structure to provide a direct discussion of what kind of structures might be effective for extracting features from EEG signals.
We apply repeat experiments on both mixed-subject search and subject-specific search and selected searched structure with top 8 highest testing accuracy in Figure~\ref{fig:vis}. 
In both scenarios, we could postulate that parallel structure takes a major partition.
We think this result is reasonable since operators arranged in parallel formation could extract the input feature with multiple scales, which means the output node of each cell could apply feature fusion from parallel operators. 
It has been widely mentioned in previous neural network papers~\cite{huang2017densely,gao2019res2net,fang2020densely} that feature fusion from various scales could improve the accuracy. 
According to our searched structures, we believe this conclusion also holds for EEG signals, especially fusion between channel-wise operators ($1\times n$) and time-wise operators ($n\times 1$). 
Meanwhile, another observation is that the structures from subject-specific search provide more diversity than those from a mixed subjects search.
This phenomenon is caused by the subject difference EEG signals.

\begin{figure*}[hbpt]{
\centering
\subfigure[Visualization of Mixed Search]{
\includegraphics[width=0.9\textwidth]{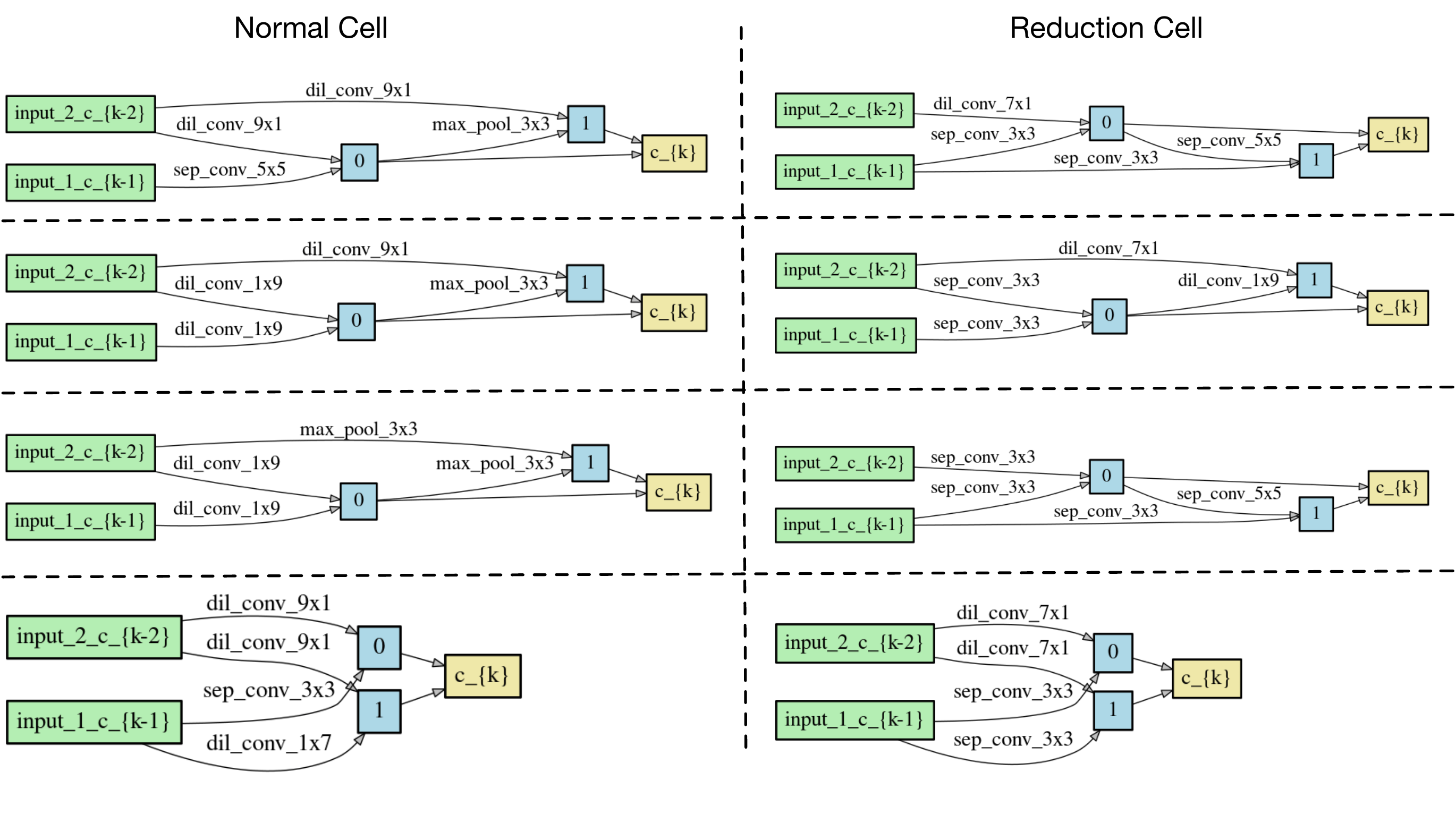}\label{fig:mixed_vis_seed}
}
\subfigure[Visualization of Subject-Specific Search]{
\includegraphics[width=0.9\textwidth]{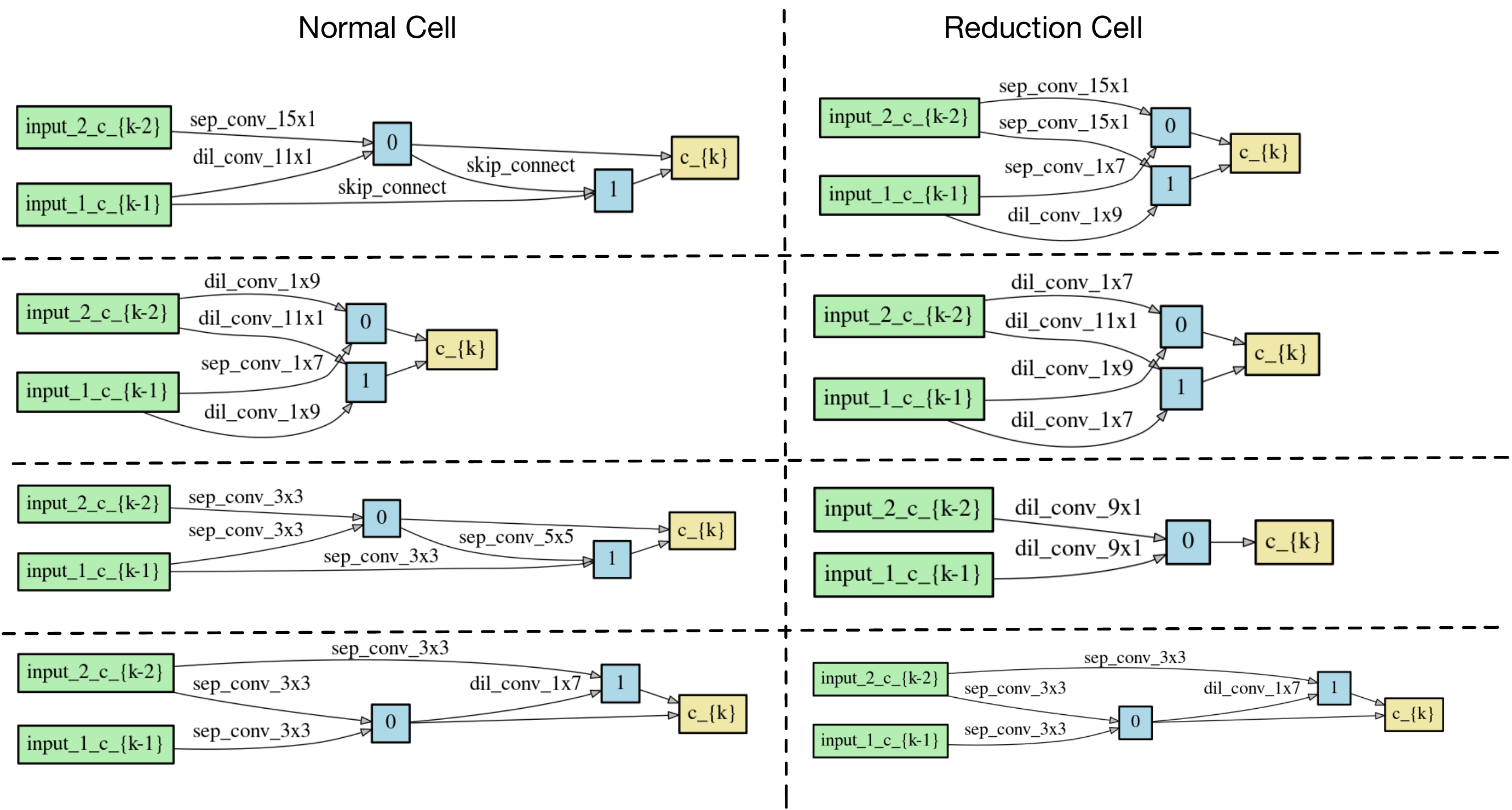}\label{fig:ss_vis_seed}
}
\caption{\label{fig:vis_seed} The searched structure on SEED V dataset, where Figure~\ref{fig:mixed_vis_seed} and ~\ref{fig:ss_vis_seed} respectively denotes the results on mixed-subject search and subject-specific search.}
}
\end{figure*}

\subsubsection{Searched Results on SEED V Datasets}
We visualize the searched structures on the SEED V dataset basically following the setting on the BCI Competition IV dataset. 
The results are visualized in Figure~\ref{fig:vis_seed}. 
It is observed that the structure searched in the SEED V dataset (Emotion) contains fewer skip connections. 
Oppositely, convolution operators with kernel sizes $3x3$ and $5x5$ are selected with a higher probability. 
This means that the Emotion tasks may need more operators to extract useful features from the raw signal. 
The classification performance supports that the Emotion tasks are normally harder compared to MI.  
In that case, we think this phenomenon is rational. 
It is also observed that the operator searched in the MI task is more `narrow' than it in Emotion tasks (eg. MI tasks normally have dilation convolution 9x1, even 17x1, however, Emotion tasks normally have separable convolution 3x3 and dilation convolution 7x1.) 
This suggests that the MI classification needs more abilities to extract long-time-wise information. 
This observation is also supported by previous works~\cite{incpet_EEGNet_paper,EEGNet_Fusion,ingolfsson2020eegtcnet}, which suggests that time-wise signal aggregation is of vital importance for MI classification. 
\clearpage

\section{Hyper Parameters}

\label{subsec:hyperparam}

To provide a reference for follow-up papers, we also conducted additional experiments to discuss which hyper-parameters could have an impact on model training for searching. 
We conduct ablation study majorly on BCI competition IV dataset for fair comparison. 

\subsubsection{Node number of the search space} 
\label{ap:nodeablation}
\begin{figure}[hbpt]
    \centering
    \includegraphics[width=0.47\textwidth]{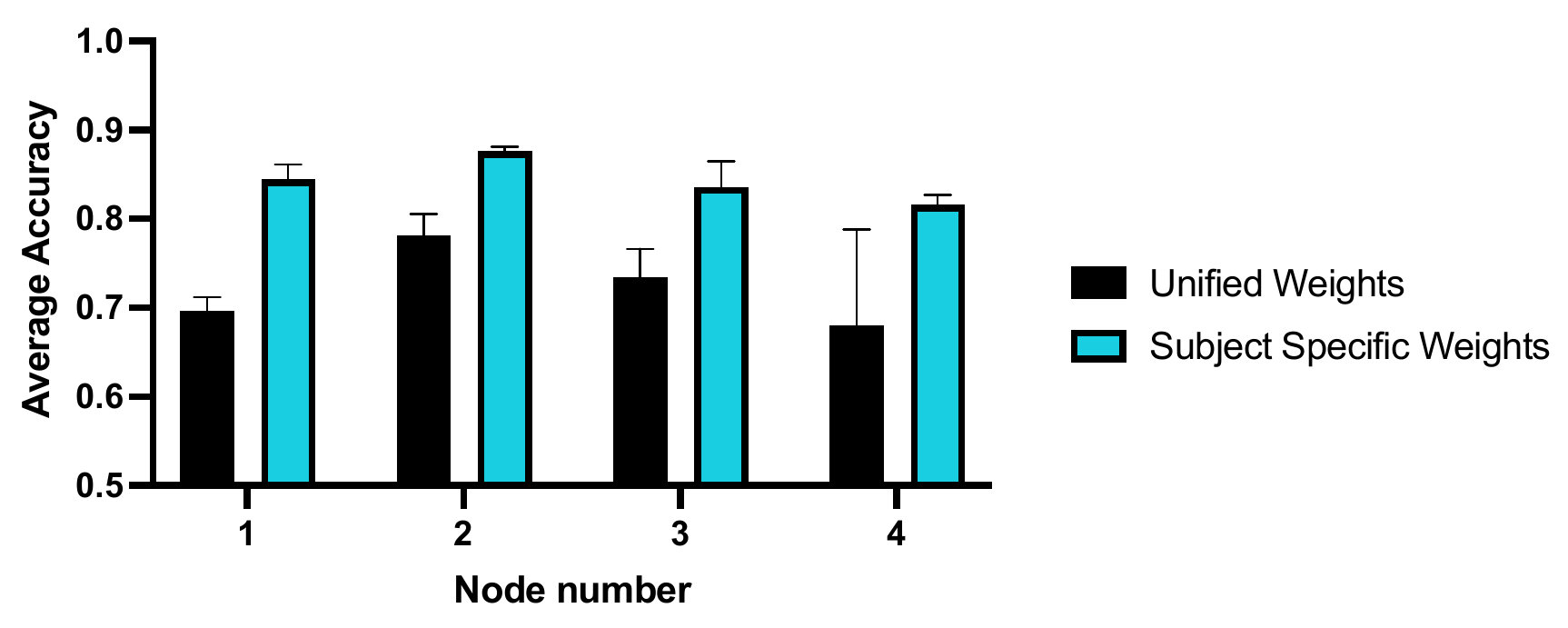}
    \caption{\label{fig:node_number}Impact of node number on model performance}
\end{figure}

As mentioned in main paper Section~\ref{sec:methods}, that the basic search-able component is called a search cell, where each search cell has two fixed input and one output nodes. Inside each search cell, the node number decides the upper bound of the edge number in it. The node number intrinsically decides the network scale of the search space. 
To give a more detailed discussion, we conduct an ablation study by only changing the node number of the search space and reporting the test accuracy in Figure~\ref{fig:node_number}. 
It is observed that the search cell with two nodes achieves the best result. 
Also, the model accuracy slightly decreases while the number of the nodes increasing. 
This phenomenon is rational as the increasing of node number lead to significantly larger search space, which is harder for searching, especially given only limited data. 
Throughout our experiments, node number 2 achieves the best result. 

\subsubsection{Batch Size}

\label{sp:batchsize}
Here, to reflect the impact of batch size for searching, we conduct an ablation study on both mixed subjects training and subject-specific training. Following other ablation studies in this paper, we selected subject 3 (S3) as representative of the subject customized structure search to keep it the same as previous results.
\begin{figure}[hbpt]
    \centering
    \includegraphics[width=0.47\textwidth]{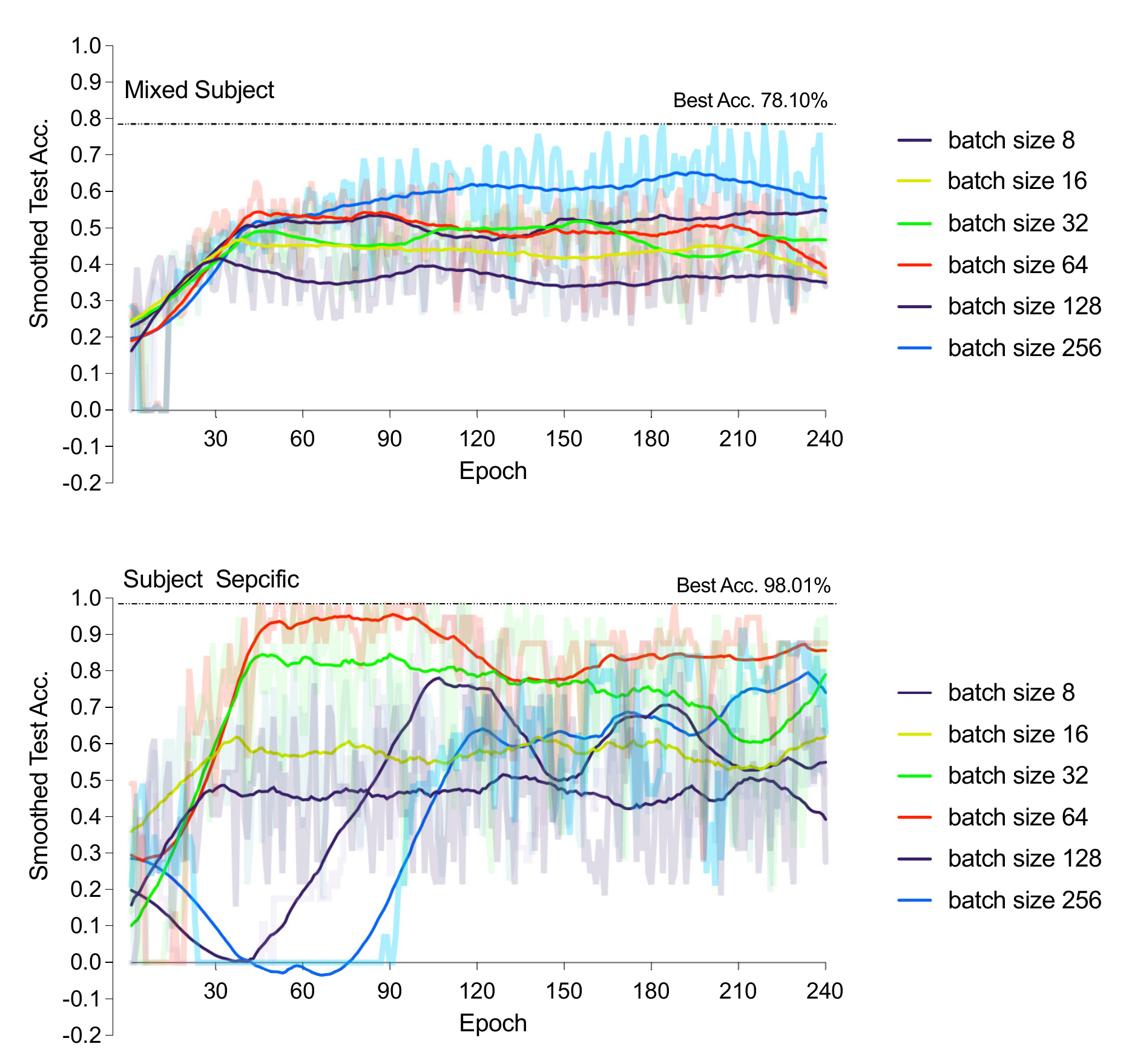}
    \caption{\label{fig:bsab}Ablation study on batch size for searching, where the upper figure report results of searching on mixed subjects and the lower figure report results of subject-specific searching.}
\end{figure}
The results are shown in Figure~\ref{fig:bsab}, where it should be noted that lines with lighter color are the actual data. 
To better reflect the trend of accuracy change, we apply the value smooth on the accuracy curves, represented as lines with clear color. For mixed search, we observe a clear trend that a larger batch size would bring a significant improvement on classification accuracy, where training with batch size 256 could reach $78.10\%$ test accuracy on all subjects. 
However, batch size 128, 64, 32 respectively reach $69.8\%$, $68.2\%$ and $65.1\%$, where the performance of batch size 16 and 8 is far below the best accuracy. 
We argue this is rational for mixed subjects search. 
The larger the batch size, the more subjects would be included in one optimization step. Including more subjects in optimization would be crucial for the model to learn common features across different subjects. Otherwise, if the batch size is too small such as 8, the model may learn biased features for only one subject, which may lead to even lower accuracy on other subjects for mixed subjects training.
For a subject-specific search, the larger a batch size is does not mean the outcomes are the better. 
Take subject 3 in Figure~\ref{fig:bsab} as an instance, the training trail with batch size 32 reaches the highest performance $98.01\%$. 
Actually, all batch sizes could reach compared good results (above $90\%$) on subject 3.
This result suggests that the batch size does not have a clear impact on the subject-specific training.
It should be also noted that, for large batch size 256 and 128, the accuracy is lower than others.
We argue this phenomenon is rational since we control the training epochs the same, larger batch size leads to smaller iteration steps. 
As for each subject, the training samples are extremely limited (500 to 800 trails). 
This may lead to an under-fit for larger batch sizes.

\subsubsection{Network Scale Constraint}
\label{ap:networkscale}
\begin{figure}
    \centering
    \includegraphics[width=0.45\textwidth]{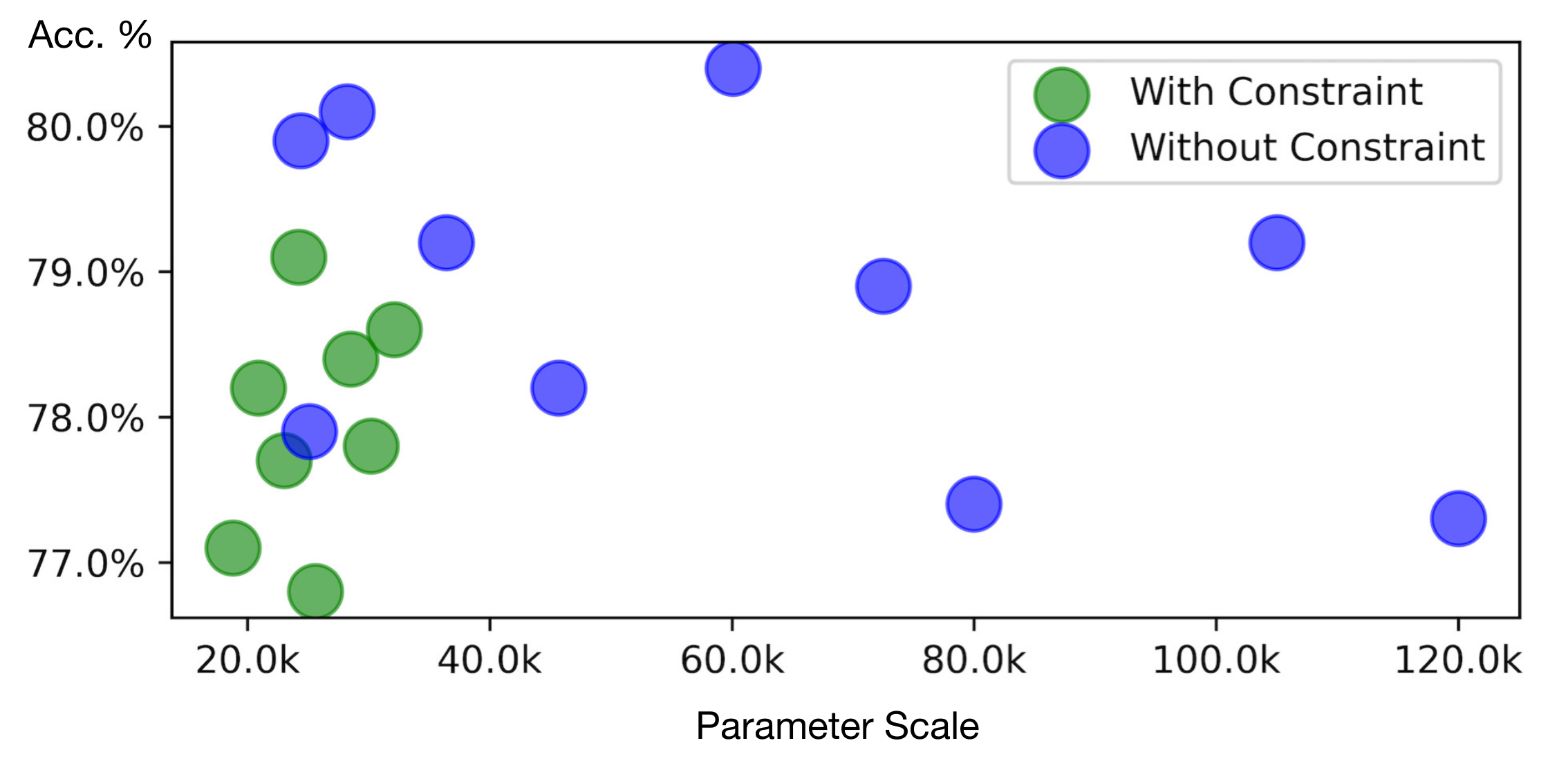}
    \caption{The parameter scale vs. performance distribution on BCI Competition IV-2a dataset, where green and blue points respectively denote the searched results from EDLAS with or without structure constraint. }
    \label{fig:ParameterScale}
\end{figure}

As we mentioned in main paper Section~\ref{subsec: constraint} that the proposed CTNAS-EEG applied network scale constraint for practical deployment for EEG tasks. 
It could be observed that, if without the structure constraint, the searched parameter scale varies from $24.1\sim120k$. 
However, according to our repeat experiments the best performance is not achieved under the largest parameter scale.
To report model scales and the performance corresponding to them.
By running repeat experiments, we report the parameter scales with performances in Figure~\ref{fig:ParameterScale}. 
It could be observed that, for searched results, the model scale is not the larger the better. 
For structure constrained search, the searched models could perform multiple levels of accuracy under the same parameter scale. 
For structure search without constraint, the models with similar accuracy may have very different parameter scales. 
However, the accuracy of a very large network scale, such as $120k$ and $105k$, suggest a clear accuracy drop compared to the best accuracy. 
We believe that it is because the large model structure might lead to more severe over-fitting on the training set. 

\subsubsection{Network Sparsity Constraint}
Except for network scale constraint in main paper Section~\ref{subsec: constraint}, CTNAS-EEG also propose to increase the probability of sampling skip connection (which could be also regarded as network sparsity) to give more instant gradient backward path to the network.
As the EEG signals are normally with a low signal-to-noise ratio, keep the network shallower could help stabilizing the training process. 
\begin{figure}[hpbt]
    \centering
    \includegraphics[width=0.45\textwidth]{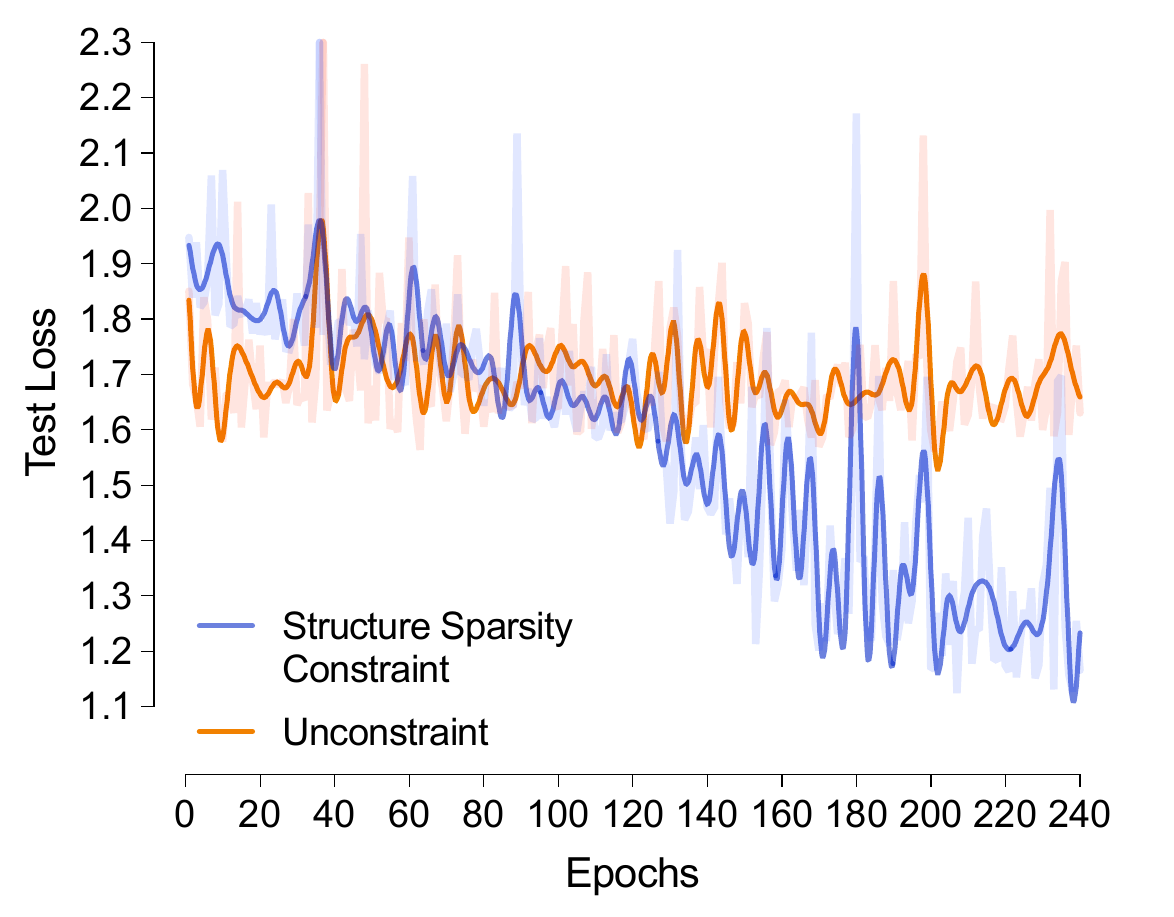}
    \caption{Comparison between sparsity constrained search and unconstrained search}
    \label{fig:sparsity constraint}
\end{figure}
Meanwhile, the CTNAS-EEG anealing the sparsity constraint along with the training process, which means that the network is free to increase capacity at later stages to while the network is converged to certain degree. 
Here we simply visualize the test loss curve during the training period in Figure~\ref{fig:sparsity constraint} to illustrate the efficiency of the proposed sparsity constraint. 
Please note that the visualization of both curve only count in classification loss for fair comparison.
It is observed that, though structure constraint may lead to slight higher classification loss at early stage, the constraint searching is compared have less fluctuation at early stage. 
Along with the constraint annealing, the constrained method (blue line) could reach test loss at later stage.
The visualization supports the efficiency of our method.

\end{document}